\documentclass[twocolumn,showpacs,superscriptaddress]{revtex4}

\usepackage{graphicx}
\usepackage{dcolumn}
\usepackage{bm}
\usepackage{multirow}

\newcommand{\eq}[1]{(\ref{#1})}
\newcommand{\matrixtop}[1]{\buildrel\leftrightarrow\over{#1}}

\newcommand{\rhobf}{\bm{\rho}}
\newcommand{\Gammabf}{\bm{\Gamma}}
\newcommand{\bun}{\hat{\mathbf{b}}}

\newcommand{\zun}{\hat{\bm{\zeta}}}
\newcommand{\matI}{\matrixtop{\mathbf{I}}}
\newcommand{\bA}{\mathbf{A}}
\newcommand{\bB}{\mathbf{B}}
\newcommand{\bE}{\mathbf{E}}
\newcommand{\bw}{\mathbf{w}}

\newcommand{\bx}{\mathbf{x}}
\newcommand{\bk}{\mathbf{k}}
\newcommand{\bX}{\mathbf{X}}
\newcommand{\bv}{\mathbf{v}}
\newcommand{\bV}{\mathbf{V}}

\newcommand{\pivar}{\textsf{P}}
\newcommand{\kvar}{\textsf{k}}
\newcommand{\wvar}{\textsf{w}}

\begin{document}


\title{Up-down symmetry of the turbulent transport of toroidal angular momentum in tokamaks}

\author{Felix I. Parra}
\affiliation{%
Rudolf Peierls Centre for Theoretical Physics, University of Oxford, Oxford, OX1 3NP, UK%
}%
\affiliation{%
Isaac Newton Institute for Mathematical Sciences, Cambridge CB3
0EH, UK
}%
\author{Michael Barnes}
\affiliation{%
Rudolf Peierls Centre for Theoretical Physics, University of Oxford, Oxford, OX1 3NP, UK%
}%
\affiliation{%
Euratom/CCFE Fusion Association, Culham Science Centre, Abingdon
OX14 3DB, UK
}%
\affiliation{%
Isaac Newton Institute for Mathematical Sciences, Cambridge CB3
0EH, UK
}%
\author{Arthur G. Peeters}%
\affiliation{%
Universit\"{a}t Bayreuth, 95440 Bayreuth, Germany
}%
\affiliation{%
Isaac Newton Institute for Mathematical Sciences, Cambridge CB3
0EH, UK
}%

\date{\today}

\begin{abstract}
Two symmetries of the local nonlinear $\delta f$ gyrokinetic
system of equations in tokamaks in the high flow regime are
presented. The turbulent transport of toroidal angular momentum
changes sign under an up-down reflection of the tokamak and a sign
change of both the rotation and the rotation shear. Thus, the
turbulent transport of toroidal angular momentum must vanish for
up-down symmetric tokamaks in the absence of both rotation and
rotation shear. This has important implications for the modeling
of spontaneous rotation.
\end{abstract}

\maketitle

\section{Introduction}

Understanding how momentum is distributed within a tokamak is
crucial to explaining experimentally observed regions of reduced
turbulent transport like Internal Transport Barriers (ITBs)
\cite{connor04} and edge pedestals \cite{connor00}. In these
regions of reduced transport, turbulence is suppressed by strong
velocity gradients, which are determined by the transport of
toroidal angular momentum. Nonlinear $\delta f$ flux tube
simulations \cite{dorland00, candy03a, dannert05, peeters09c} are
extensively used to study turbulent transport of toroidal angular
momentum in tokamaks \cite{peeters07, waltz07, camenen09, roach09,
casson09, casson10, barnes10, highcock10}. The time averaged
turbulent transport of momentum is found to be zero for up-down
symmetric tokamaks in the absence of toroidal velocity and
toroidal velocity gradient. This behavior was addressed in
Ref.~\cite{peeters05}, where the transport of parallel momentum
due to electrostatic fluctuations was studied with a quasilinear
model. Ref.~\cite{peeters05} showed that the linearized
gyrokinetic equation has a symmetry that results in the transport
of parallel momentum vanishing for up-down symmetric tokamaks in
the absence of velocity and velocity shear. Moreover, this
analysis was successfully employed to identify the up-down
asymmetry as a drive for momentum transport \cite{camenen09}.

However, there are six major effects that were not treated in the
analysis of Ref.~\cite{peeters05}. First, the symmetry given in
Ref.~\cite{peeters05} is only valid for the linearized equation.
Second, the formulation was collisionless. Third, the effect of
the $\bE \times \bB$ shear on transport of momentum
\cite{dominguez93, roach09, casson09} was not considered. Fourth,
the Mach number was assumed to be small and the effect of the
centrifugal force \cite{casson10} was neglected. Fifth, in the
evaluation of the radial transport of toroidal angular momentum,
the contribution of the component of the velocity perpendicular to
the magnetic field was neglected. Sixth, electromagnetic effects
were not taken into account. In this article, we present the
complete symmetry argument. We consider the high flow regime, for
which the mean velocity is comparable to the thermal speed of the
ions, and we use the local approximation, for which the turbulent
transport only depends on local quantities. In this limit, the
sign of the radial transport of toroidal angular momentum changes
under a sign change of the average velocity and its gradient and
an up-down reflection. Thus, for an up-down symmetric tokamak
without any velocity or velocity shear, the momentum transport is
zero.

The symmetry presented here is broken if higher order terms in the
gyrokinetic expansion in $\epsilon_i = \rho_i/a \ll 1$ are kept,
giving transport of momentum that is small in $\epsilon_i$
compared to the high flow results. Here $\rho_i$ is the ion
gyroradius and $a$ is the minor radius of the tokamak. The effect
of these higher order terms on transport of momentum is discussed
in Ref.~\cite{parra10a}, where a higher order equation valid in
the limit of small poloidal magnetic field is proposed. The
complete higher order gyrokinetic Fokker-Planck equation valid for
any magnetic geometry was self-consistently calculated for the
first time in Ref.~\cite{parra11a} for the electrostatic
approximation.

The remainder of this article is organized as follows. We first
present the local $\delta f$ gyrokinetic model in the high flow
regime in section~\ref{sec:model}. We then show in
section~\ref{sec:symmetry} that the gyrokinetic equations have two
fundamental symmetries that involve an up-down reflection.
Finally, the consequences of these symmetries for transport of
momentum are discussed in section~\ref{sec:conclusion}. The most
cumbersome parts of the calculation are given in Appendices
\ref{app:collision}-\ref{app:updown2}.

\section{Local $\delta f$ gyrokinetic formulation} \label{sec:model}

In this section we present the local $\delta f$ gyrokinetic system
of equations in the high flow ordering, for which the average
velocity of the ions $\bV_i$ is of the order of the ion thermal
speed $v_{ti}$.

First, in subsection~\ref{sub:equilibrium}, we review the
magnetohydrodynamic (MHD) equilibrium for a tokamak and we discuss
the background density, temperature and velocity. We then present
in subsection~\ref{sub:fluctuations} the system of equations for
the fluctuating pieces of the distribution function and the
electromagnetic field. These equations are simplified in the local
limit in subsection~\ref{sub:local}. We finish by giving the
radial flux of toroidal angular momentum as a function of the
fluctuating distribution function and electromagnetic fields in
subsection~\ref{sub:fluxes}.

\subsection{Equilibrium} \label{sub:equilibrium}

The background axisymmetric magnetic field of a tokamak is given
by
\begin{equation} \label{defB}
\bB = I \nabla \zeta + \nabla \zeta \times \nabla \psi,
\end{equation}
where $\psi$ is the poloidal magnetic flux, $\zeta$ is the
toroidal angle, $\nabla \zeta = \zun/R$, with $\zun$ the unit
vector in the toroidal direction and $R$ the major radius, and $I
= RB_\zeta$, with $B_\zeta$ the toroidal magnetic field. We use
$\psi$, $\zeta$ and a poloidal angle $\theta$ as our coordinates.
The Jacobian of the transformation from cartesian coordinates
$\bx$ into $\psi$, $\theta$ and $\zeta$ is
\begin{equation} \label{jacob}
\mathcal{J} = |(\nabla \psi\times \nabla \theta) \cdot \nabla
\zeta|^{-1} = |\bB \cdot \nabla \theta|^{-1},
\end{equation}
where we take the absolute value to avoid having a sign in the
Jacobian.

In the high flow ordering, the electric field is electrostatic and
purely radial to lowest order, and the potential is given by
$e\Phi_{-1}(\psi, t)/T_e \sim \epsilon_i^{-1}$, with $T_e$ the
electron temperature and $e$ the magnitude of the electron charge.
To lowest order, all species rotate at the same speed within a
flux surface, i.e., $\bV_s = \bV_i = R\Omega_\zeta (\psi, t) \zun
\sim v_{ti}$ \cite{hinton85, connor87, catto87}, where the
subscript $s$ indicates the species, and $\Omega_\zeta (\psi, t) =
- c (\partial \Phi_{-1}/\partial \psi)$ is the rotation frequency.
To next order, there is an axisymmetric component of the
potential, $e\Phi_0 (\psi, \theta, t)/T_e \sim 1$ that ensures
quasineutrality within a flux surface.

It is convenient to solve the equations in the frame rotating with
$\Omega_\zeta$. Thus, from now on we use the peculiar velocity
$\bw = \bv - R\Omega_\zeta \zun$, where $\bv$ is the velocity in
the laboratory frame. In the frame rotating with $\Omega_\zeta$,
the lowest order distribution functions are stationary Maxwellians
$f_{Ms} = n_s (m_s/2\pi T_s)^{3/2} \exp ( - m_s w^2/2T_s)$, where
$m_s$ is the species mass, and $n_s$ and $T_s$ are the species
density and temperature, respectively. The temperatures $T_s
(\psi, t)$ are flux functions. Due to the centrifugal force and
the electrostatic potential, the densities are not constant in a
flux surface. Instead, they are given by
\begin{equation} \label{neq}
n_s (\psi, \theta, t) = \eta_s (\psi, t) \exp \left ( - \frac{Z_s
e \Phi_0}{T_s} + \frac{m_s \Omega_\zeta^2 R^2}{2T_s} \right ),
\end{equation}
with $\eta_s (\psi, t)$ a flux function and $Z_s e$ the species
charge. Imposing quasineutrality,
\begin{equation}
\sum_s Z_s e n_s = 0,
\end{equation}
gives the piece of the electrostatic potential $\Phi_0 (\psi,
\theta, t)$.

With these lowest order equilibrium distribution functions,
$I(\psi, t) = R B_\zeta$ is a flux function, and the poloidal
component of the magnetic field is determined by the
Grad-Shafranov equation
\begin{equation}
R^2 \nabla \cdot \left ( \frac{\nabla \psi}{R^2} \right ) = - I
\frac{\partial I}{\partial \psi} - 4\pi R^2 \left. \frac{\partial
p}{\partial \psi} \right |_R,
\end{equation}
where the pressure is
\begin{equation}
p(\psi, \theta, t) = \sum_s \eta_s T_s \exp \left ( - \frac{Z_s e
\Phi_0}{T_s} + \frac{m_s \Omega_\zeta^2 R^2}{2T_s} \right )
\end{equation}
and
\begin{equation}
\left. \frac{\partial p}{\partial \psi} \right |_R =
\frac{\partial p}{\partial \psi} - \sum_s n_s m_s \Omega_\zeta^2 R
\frac{\partial R}{\partial \psi}.
\end{equation}

\subsection{Equations for the turbulent fluctuations} \label{sub:fluctuations}

The $\delta f$ formulation in the high flow regime is discussed in
\cite{artun94, sugama98, peeters09a, abel11}. It determines the
turbulent pieces of the distribution function, $\delta \bar f_s
(\bx, \bw, t) \sim \epsilon_i f_{Ms}$, the electrostatic
potential, $\bar \phi (\bx, t) \sim \epsilon_i T_e/e$, the
parallel component of the vector potential, $\bar A_{||} (\bx, t)
\sim \epsilon_i \rho_i B$, and the parallel component of the
magnetic field, $\bar B_{||} (\bx, t) \sim \epsilon_i B$. The
perturbed electric and magnetic fields in the frame rotating with
velocity $\Omega_\zeta$ are to lowest order $\delta \bar \bE = -
\nabla_\bot \bar \phi$ and $\delta \bar \bB = \nabla_\bot \bar
A_{||} \times \bun + \bar B_{||} \bun$, where $\bun = \bB/B$ is
the unit vector in the direction of the background magnetic field,
and $\nabla_\bot = \nabla - \bun \bun \cdot \nabla$ is the
gradient perpendicular to the background magnetic field. We assume
that the perpendicular gradient of the fluctuations is large, on
the order of the inverse of the ion gyroradius, $\nabla_\bot \sim
1/\rho_i$, whereas the parallel gradient is comparable to the
inverse of the characteristic length of the device, $\bun \cdot
\nabla \sim 1/a$. The overline $\,\bar{\,}\,$ over the perturbed
quantities will be useful later because the quantities without
this overline will be their Fourier transforms.

To eliminate the fast gyrofrequency time scale, it is necessary to
change to gyrokinetic phase space coordinates. We use the guiding
center position $\bX = \bx + \Omega_s^{-1} \bw \times \bun$, the
velocity parallel to the magnetic field $w_{||} = \bw \cdot \bun$,
the magnetic moment $\mu = w_\bot^2/2B$ and the gyrophase $\varphi
= \arctan [(\bw \times \bun) \cdot \nabla \psi / (\bw \cdot \nabla
\psi) ]$, where $\Omega_s = Z_s e B/m_s c$ is the species
gyrofrequency and $\bw_\bot = \bw - w_{||} \bun$ is the velocity
perpendicular to the background magnetic field. With these new
coordinates, it is easy to average out the gyrofrequency time
scale. Extracting the Maxwell-Boltzmann response from the
distribution function, i.e., $\delta \bar f_s = \bar h_s - (Z_s e
\bar \phi/T_s) f_{Ms}$, $\bar h_s (\bX, w_{||}, \mu, t)$ can be
proven to be gyrophase independent, and the $\delta f$ kinetic
equation becomes
\begin{eqnarray} \label{kinetic}
\frac{\partial \bar h_s}{\partial t} + R \Omega_\zeta \zun \cdot
\nabla_\bX \bar h_s + w_{||} \bun \cdot \nabla_\bX \bar h_s +
\bv_{d, s} \cdot \nabla_\bX \bar h_s \nonumber \\ + a_{||, s}
\frac{\partial \bar h_s}{\partial w_{||}} - \sum_{s^\prime} \left
\langle \bar C_{ss^\prime}^{(\ell)} \right \rangle_s + \bar
\bv_{\chi,s} \cdot \nabla_\bX \bar h_s \nonumber \\= \frac{Z_s e
f_{Ms}}{T_s} \left ( \frac{\partial \langle \bar \chi
\rangle_s}{\partial t} + R \Omega_\zeta \zun
\cdot \nabla_\bX \langle \bar \chi \rangle_s \right ) \nonumber\\
- \bar \bv_{\chi,s} \cdot \Bigg [ \nabla_\bX f_{Ms} + \frac{m_s
f_{Ms}}{T_s} \bigg ( \frac{Iw_{||}}{B} \nabla_\bX \Omega_\zeta
\nonumber\\+ \frac{Z_s e}{m_s} \nabla_\bX \Phi_0 - R\Omega_\zeta^2
\nabla_\bX R \bigg ) \Bigg ],
\end{eqnarray}
where the Maxwellian is written in the new variables as $f_{Ms} =
n_s (m_s/2\pi T_s)^{3/2} \exp [ - m_s (w_{||}^2/2 + \mu B)/T_s ]$.
The drift
\begin{equation} \label{vchi}
\bar \bv_{\chi,s} = - \frac{c}{B} \nabla_\bX \langle \bar \chi
\rangle_s \times \bun
\end{equation}
accounts for the parallel motion along the perturbed magnetic
field lines and for the turbulent $\bE \times \bB$ and $\nabla B$
drifts. The generalized potential $\langle \bar \chi \rangle_s$ is
given by
\begin{eqnarray} \label{chi}
\langle \bar \chi \rangle_s = \langle \bar \phi (\bX + \rhobf_s,
t) \rangle_s - \frac{1}{c} w_{||} \langle \bar A_{||} (\bX +
\rhobf_s, t) \rangle_s \nonumber \\ - \frac{1}{c} \langle \bw_\bot
\cdot \bar \bA_\bot (\bX + \rhobf_s, t) \rangle_s,
\end{eqnarray}
where $\bar \bA_\bot$ is related to $\bar B_{||}$ by $\bar B_{||}
= \bun \cdot (\nabla \times \bar \bA_\bot)$ and $\langle \ldots
\rangle_s$ is the gyroaverage holding $\bX$, $w_{||}$, $\mu$ and
$t$ fixed. The dependence of $\bar \phi$, $\bar A_{||}$ and $\bar
\bA_\bot$ on the gyrophase is through $\bx = \bX + \rhobf_s$ that
contains the gyroradius $\rhobf_s = - \Omega_s^{-1} \bw \times
\bun$, also given by
\begin{eqnarray} \label{rho}
\rhobf_s (\mu, \varphi) = \frac{\sqrt{2\mu B}}{\Omega_s |\nabla
\psi|} [ - \sin \varphi\, \nabla \psi + \cos \varphi\, (\bun
\times \nabla \psi ) ].
\end{eqnarray}
The perpendicular drifts $\bv_{d,s}$ are
\begin{equation} \label{vd}
\bv_{d,s} = \bv_{E0} + \bv_{M,s} + \bv_{co,s} + \bv_{cf,s},
\end{equation}
where $\bv_{E0} = - (c/B) \nabla_\bX \Phi_0 \times \bun$ is the
$\bE \times \bB$ drift due to the background axisymmetric
potential $\Phi_0$, $\bv_{M,s} = (\mu/\Omega_s) \bun \times
\nabla_\bX B + (w_{||}^2/\Omega_s) \bun \times ( \bun \cdot
\nabla_\bX \bun)$ are the $\nabla B$ and curvature drifts,
$\bv_{co,s} = (2 w_{||} \Omega_\zeta/\Omega_s) \bun \times
[(\nabla_\bX R \times \zun) \times \bun]$ is the Coriolis drift,
and $\bv_{cf,s} = - (R \Omega_\zeta^2/\Omega_s) \bun \times
\nabla_\bX R$ is the centrifugal force drift. The parallel
acceleration is
\begin{equation} \label{apar}
a_{||,s} = - \mu \bun \cdot \nabla_\bX B - \frac{Z_s e}{m_s} \bun
\cdot \nabla_\bX \Phi_0 + R \Omega_\zeta^2 \bun \cdot \nabla_\bX
R.
\end{equation}
Finally, $\langle \bar C^{(\ell)}_{ss^\prime} \rangle_s = \langle
C_{ss^\prime} [ \bar h_s (\bx - \rhobf_s, w_{||}, \mu, t),
f_{Ms^\prime} ] \rangle_s + \langle C_{ss^\prime} [ f_{Ms}, \bar
h_{s^\prime} (\bx - \rhobf_{s^\prime}, w_{||}, \mu, t)] \rangle_s$
is the gyroaveraged linearized collision operator between species
$s$ and $s^\prime$. Here $C_{ss^\prime} [ f_s, f_{s^\prime} ]$ is
the bilinear full collision operator, and we have emphasized that
$\bar h_s (\bX, w_{||}, \mu, t)$ and $\bar h_{s^\prime} (\bX,
w_{||}, \mu, t)$ depend on velocity space not only through the
variables $w_{||}$ and $\mu$ but also through $\bX = \bx -
\rhobf_s (\mu, \varphi)$. The linearized collision operator is
linear in both $\bar h_s$ and $\bar h_{s^\prime}$. To show the
symmetries of the local $\delta f$ gyrokinetic model, it is
convenient to write the collision operator using the Rosenbluth
potentials \cite{rosenbluth57}. The particular expressions that we
use are given in Appendix~\ref{app:collision}.

The fluctuations of the electrostatic potential are obtained using
the quasineutrality equation, $\sum_s Z_s \int d^3w\, \delta \bar
f_s = 0$, where the integral in velocity space is taken at fixed
$\bx$. Employing $d^3w = B\, dw_{||}\, d\mu\, d\varphi$, we find
that quasineutrality becomes
\begin{equation} \label{qn}
\sum_s Z_s B \int dw_{||}\, d\mu\, d\varphi\, \bar h_s(\bx -
\rhobf_s) - \sum_s \frac{Z_s^2 e n_s \bar \phi}{T_s} = 0,
\end{equation}
where we have stressed that $\bar h_s$ depends on velocity not
only through $w_{||}$ and $\mu$, but also through $\bX = \bx -
\rhobf_s (\mu, \varphi)$. The parallel vector potential $\bar
A_{||}$ and the parallel magnetic field $\bar B_{||}$ are both
determined by the current equation $\nabla \times ( \nabla \bar
A_{||} \times \bun + \bar B_{||} \bun ) = (4\pi/c) \sum_s Z_s e
\int d^3w\, \delta \bar f_s \bw$. The parallel component of this
equation gives
\begin{equation} \label{parcurr}
- \nabla_\bot^2 \bar A_{||} = \sum_s \frac{4\pi Z_s e B}{c} \int
dw_{||}\, d\mu\, d\varphi\, \bar h_s(\bx - \rhobf_s) w_{||},
\end{equation}
and the perpendicular component is
\begin{equation} \label{perppress}
\nabla \bar B_{||} \times \bun = \sum_s \frac{4\pi Z_s e B}{c}
\int dw_{||}\, d\mu\, d\varphi\, \bar h_s (\bx - \rhobf_s)
\bw_\bot.
\end{equation}
We have neglected the parallel gradient $\bun \cdot \nabla \sim
1/a$ of $\bar A_{||}$ and the gradient $\nabla \sim 1/a$ of the
background magnetic field $\bB$ because they are much smaller than
the perpendicular gradients $\nabla_\bot \sim 1/\rho_i$ of the
fluctuations.

\subsection{Local approximation} \label{sub:local}

Since the characteristic perpendicular length of the turbulent
structures is much shorter than the parallel length, it is
convenient to describe the turbulent pieces $\bar h_s (\psi(\bX),
\alpha(\bX, t), \theta(\bX), w_{||}, \mu, t)$, $\bar \phi
(\psi(\bx), \alpha(\bx, t), \theta(\bx), t)$, $\bar A_{||}
(\psi(\bx), \alpha(\bx, t), \theta(\bx), t)$ and $\bar B_{||}
(\psi(\bx), \alpha(\bx, t), \theta(\bx), t)$ by two coordinates
perpendicular to the magnetic field line, $\psi$ and $\alpha$, and
a coordinate to locate the position along the magnetic field that
in this case is the poloidal angle $\theta$. We stress that the
distribution function depends on the guiding center position $\bX$
and that the electromagnetic fields depend on the position $\bx$.
Here $\alpha (\bx, t)$ is the coordinate in the direction
perpendicular to the magnetic field within the flux surface and
rotating with toroidal angular velocity $\Omega_\zeta$, defined
such that $\bB = \nabla \alpha \times \nabla \psi$ and $\partial
\alpha/\partial t + \Omega_\zeta (\partial \alpha/\partial \zeta)
= 0$, i.e.,
\begin{equation} \label{alpha}
\alpha = \zeta - q(\psi) \, \vartheta(\psi, \theta) -
\Omega_\zeta(\psi) t,
\end{equation}
where
\begin{equation} \label{vartheta}
\vartheta (\psi, \theta) = \frac{I(\psi)}{q(\psi)} \int_0^\theta
d\theta^\prime\, (R^2 \bB \cdot \nabla \theta)^{-1} |_{\psi,
\theta^\prime}
\end{equation}
and $q(\psi) = |(I/2\pi) \int_0^{2\pi} d\theta\, (R^2 \bB \cdot
\nabla \theta)^{-1}|$ is the safety factor. The advantage of using
the time dependent variable $\alpha$ is that the $\bE \times \bB$
shear \cite{casson09, dominguez93} is included in the formulation
in a natural way, as we will see.

In the local approximation, the characteristic perpendicular size
of the turbulent structures is assumed to be small compared to the
background length scale. The local $\delta f$ equations only
depend on $\psi$ and $\alpha$ through the unknowns $\bar h_s$,
$\bar \phi$, $\bar A_{||}$ and $\bar B_{||}$. It is then more
convenient to Fourier analyze the distribution function and the
electromagnetic fields. Defining $g = (\Delta \psi \, \Delta
\alpha)^{-1} \int_{\Delta \psi} d\psi \int_{\Delta \alpha}
d\alpha\, \bar g \exp( - ik_\psi \psi - ik_\alpha \alpha )$, with
$i = \sqrt{-1}$, $g = \{ h_s, \phi, A_{||}, B_{||} \}$, $\bar g =
\{ \bar h_s, \bar \phi, \bar A_{||}, \bar B_{||} \}$, and $\Delta
\psi$ and $\Delta \alpha$ the size of the domain of interest, and
using Appendix~\ref{app:kspace}, the $\delta f$ gyrokinetic
equation becomes
\begin{eqnarray} \label{kinetickspace}
\frac{\partial h_s}{\partial t} + w_{||} \bun \cdot \nabla \theta
\frac{\partial h_s}{\partial \theta} + i k_\psi h_s v_{d, s}^\psi
+ i k_\alpha h_s v_{d, s}^\alpha \nonumber \\ + a_{||, s}
\frac{\partial h_s}{\partial w_{||}} - \sum_{s^\prime} \left
\langle C_{ss^\prime}^{(\ell)} \right \rangle_s + \{ \langle \chi
\rangle_s, h_s \} \nonumber \\ = \frac{Z_s e f_{Ms}}{T_s}
\frac{\partial \langle \chi \rangle_s}{\partial t} - f_{Ms}
v_{\chi, s}^\psi \Bigg [ \frac{1}{n_s} \frac{\partial
n_s}{\partial \psi} + \frac{m_s Iw_{||}}{BT_s} \frac{\partial
\Omega_\zeta}{\partial \psi} \nonumber\\ + \frac{Z_s e}{T_s}
\frac{\partial \Phi_0}{\partial \psi} - \frac{m_s R
\Omega_\zeta^2}{T_s} \frac{\partial R}{\partial \psi} + \bigg (
\frac{m_s w^2}{2T_s} - \frac{3}{2} \bigg ) \frac{1}{T_s}
\frac{\partial T_s}{\partial \psi} \Bigg ].
\end{eqnarray}
Here, the turbulent components of the electromagnetic field are
contained in
\begin{equation} \label{chikspace}
\langle \chi \rangle_s = J_0 (z_s) \left ( \phi - \frac{1}{c}
w_{||} A_{||} \right ) + \frac{2 J_1 (z_s)}{z_s} \frac{m_s
\mu}{Z_s e} B_{||},
\end{equation}
where $J_n (z_s)$ is the $n$-th Bessel function of the first kind.
The function $z_s(k_\psi, k_\alpha, \theta, \mu, t)$ is
\begin{equation} \label{defz}
z_s(k_\psi, k_\alpha, \theta, \mu, t) = \frac{k_\bot \sqrt{2\mu
B}}{\Omega_s},
\end{equation}
with
\begin{equation}
k_\bot = (k_\psi^2 |\nabla \psi|^2 + 2 k_\psi k_\alpha \nabla \psi
\cdot \nabla \alpha + k_\alpha^2 |\nabla \alpha|^2)^{1/2}
\end{equation}
the magnitude of the perpendicular wavenumber. Note that the
perpendicular wavenumber increases with time because $\nabla
\alpha = \ldots - \nabla \psi (\partial \Omega_\zeta/\partial
\psi) t$. This is equivalent to the procedure generally employed
in local $\delta f$ simulations to account for the shear in the
$\bE \times \bB$ drift \cite{roach09, casson09, hammett06}. The
generalized potential $\langle \chi \rangle_s$ appears in the
radial turbulent drift
\begin{equation}
v_{\chi, s}^\psi = i k_\alpha c  \langle \chi \rangle_s
\end{equation}
and in the nonlinear term
\begin{eqnarray}
\{ \langle \chi \rangle_s, h_s \} = c \sum_{k_\psi^\prime,
k_\alpha^\prime} (k_\psi^\prime k_\alpha - k_\psi k_\alpha^\prime)
\langle \chi \rangle_s (k_\psi^\prime, k_\alpha^\prime)\nonumber\\
\times h_s (k_\psi - k_\psi^\prime, k_\alpha - k_\alpha^\prime).
\end{eqnarray}
The $\psi$ and $\alpha$ components of the drift velocity are
\begin{eqnarray} \label{vdpsikspace}
v_{d,s}^\psi = \Bigg [- \frac{cI}{B} \frac{\partial
\Phi_0}{\partial \theta} - \frac{I( w_{||}^2 + \mu B) }{B\Omega_s}
\frac{\partial B}{\partial \theta} \nonumber \\ + \frac{2 B R
\Omega_\zeta w_{||}}{\Omega_s} \frac{\partial R}{\partial \theta}
+ \frac{I R \Omega_\zeta^2}{\Omega_s} \frac{\partial R}{\partial
\theta} \Bigg ] \bun \cdot \nabla \theta
\end{eqnarray}
and
\begin{eqnarray} \label{vdalphakspace}
v_{d,s}^\alpha = - c \left [ \frac{\partial \Phi_0}{\partial \psi}
- \frac{\partial \Phi_0}{\partial \theta} \frac{(\nabla \alpha
\times \bun) \cdot \nabla \theta}{B} \right ] \nonumber \\ -
\frac{w_{||}^2 + \mu B}{\Omega_s} \left [ \frac{\partial
B}{\partial \psi} - \frac{\partial B}{\partial \theta}
\frac{(\nabla \alpha \times \bun) \cdot \nabla \theta}{B} \right ]
\nonumber \\ - \frac{4 \pi w_{||}^2}{B \Omega_s} \left.
\frac{\partial p}{\partial \psi} \right |_R + \frac{2 \Omega_\zeta
w_{||}}{\Omega_s} (\nabla R \times \zun) \cdot \nabla \alpha
\nonumber \\ + \frac{m_s c R \Omega_\zeta^2}{Z_s e} \left [
\frac{\partial R}{\partial \psi} - \frac{\partial R}{\partial
\theta} \frac{(\nabla \alpha \times \bun) \cdot \nabla \theta}{B}
\right ].
\end{eqnarray}
The parallel acceleration is
\begin{equation}
a_{||, s} = \left ( - \mu \frac{\partial B}{\partial \theta} -
\frac{Z_s e}{m_s} \frac{\partial \Phi_0}{\partial \theta} + R
\Omega_\zeta^2 \frac{\partial R}{\partial \theta} \right ) \bun
\cdot \nabla \theta.
\end{equation}
The Fourier transform of the gyroaveraged collision operator for
collisions between species $s$ and $s^\prime$, $\langle
C_{ss^\prime}^{(\ell)} \rangle_s$, is given in
Appendix~\ref{app:collisionave}.

The local versions of the quasineutrality equation \eq{qn}, the
parallel current equation \eq{parcurr} and the perpendicular
current equation \eq{perppress} are obtained in
Appendix~\ref{app:kspace2}, and are given by
\begin{equation} \label{qnkspace}
\sum_s 2 \pi Z_s B \int dw_{||}\, d\mu\, h_s J_0 (z_s) - \sum_s
\frac{Z_s^2 e n_s \phi}{T_s} = 0,
\end{equation}
\begin{equation} \label{parcurrkspace}
k_\bot^2 A_{||} = \sum_s \frac{8\pi^2 Z_s e B}{c} \int dw_{||}\,
d\mu\, h_s w_{||} J_0(z_s)
\end{equation}
and
\begin{equation} \label{perppresskspace}
\frac{B B_{||}}{4\pi} + \sum_s 2\pi m_s B^2 \int dw_{||}\, d\mu\,
h_s \mu \frac{2J_1(z_s)}{z_s} = 0.
\end{equation}
The perpendicular current equation \eq{perppresskspace} has been
rewritten as a perpendicular pressure balance.

\subsection{Radial flux of toroidal angular momentum} \label{sub:fluxes}

The radial flux of toroidal angular momentum is given by
\begin{equation}
\Pi = \sum_s \langle \Pi_s \rangle_{\Delta t} + \langle \Pi_B
\rangle_{\Delta t},
\end{equation}
where
\begin{equation} \label{eq:pis1}
\Pi_s = m_s \left \langle \left \langle R \int d^3w\, f_s (\bv
\cdot \zun) (\bv \cdot \nabla \psi) \right \rangle_\psi \right
\rangle_{\Delta \psi}
\end{equation}
is the momentum flux due to the species $s$, and
\begin{equation} \label{eq:piB1}
\Pi_B = - \frac{1}{4\pi} \left \langle \left \langle R [ (\bB +
\delta \bar{\bB}) \cdot \zun ] [ (\bB + \delta \bar{\bB}) \cdot
\nabla \psi ] \right \rangle_\psi \right \rangle_{\Delta \psi}
\end{equation}
is the momentum flux due to the Maxwell stress. Here, $\langle
\ldots \rangle_\psi = (V^\prime)^{-1} \int d\theta\, d\zeta\,
\mathcal{J} (...)$ is the flux surface average, $V^\prime \equiv
dV/d\psi = 2\pi \int d\theta\, \mathcal{J}$, and $\langle \ldots
\rangle_\mathrm{\Delta \psi} = \Delta \psi^{-1} \int_{\Delta \psi}
d\psi\, (\ldots)$ and $\langle \ldots \rangle_\mathrm{\Delta t} =
\Delta t^{-1} \int_{\Delta t} dt \, (\ldots)$ are coarse grain
averages over a radial region $\Delta \psi$ and over a time
interval $\Delta t$ larger than the decorrelation length and time
of the turbulence, respectively. In the local approximation,
$\Delta \psi$ is much shorter than the characteristic length of
variation of the background profiles of density, temperature and
velocity, and $\Delta t$ is much shorter than the transport time
scale. We remind the reader that the velocity $\bv = \bw + R
\Omega_\zeta \zun$, used in equation \eq{eq:pis1} and in some
other equations in this section, is the velocity in the laboratory
frame. Note also that the distribution function $f_s$ includes the
lowest order Maxwellian $f_{Ms}$, the turbulent contribution
$\delta \bar f_s$ and neoclassical higher order pieces. Finally,
it is worth pointing out that our definition of radial flux of
toroidal angular momentum does not have units of
$\mathrm{pressure} \times \mathrm{length}$, but of
$\mathrm{pressure} \times (\mathrm{length})^2 \times
(\mathrm{magnetic\; field})$ because our radial variable $\psi$,
which appears in the gradient $\nabla \psi$, has dimensions of
$(\mathrm{length})^2 \times (\mathrm{magnetic\; field})$.

To calculate an expression for the turbulent contribution to
$\Pi_s$, it is convenient to use the full Fokker-Planck equation
\begin{eqnarray} \label{FPequation}
\frac{\partial f_s}{\partial t} + \bv \cdot \nabla f_s + \Bigg \{
\Omega_s \bv \times \bun + \frac{Z_s e}{m_s} \bigg [ - \nabla (
\Phi_{-1} + \Phi_0) \nonumber\\ + \delta \bar \bE + \frac{1}{c}
\bw \times \delta \bar \bB \bigg ] \Bigg \} \cdot \nabla_v f_s =
\sum_{s^\prime} C_{ss^\prime} [ f_s, f_{s^\prime} ].
\end{eqnarray}
The turbulent Lorentz force is expressed in terms of the peculiar
velocity $\bw = \bv - R\Omega_\zeta \zun$ because the fluctuating
magnetic fields are calculated in the rotating frame. By taking
moments of \eq{FPequation} we obtain a simplified expression for
$\Pi_s$ \cite{parra10a, simakov07}. The $m_s R^2 (\bv \cdot
\zun)^2$ moment gives
\begin{widetext}
\begin{eqnarray}
\frac{2Z_s e R}{c} \int d^3w\, f_s (\bv \cdot \zun) (\bv \cdot
\nabla \psi) = - 2 Z_s e R^2 \int d^3w\, f_s (\bv \cdot \zun)
\Bigg ( \delta \bar \bE + \frac{1}{c} \bw \times \delta \bar \bB
\Bigg ) \cdot \zun + \frac{\partial}{\partial t}\left
[ m_s R^2 \int d^3w\, f_s (\bv \cdot \zun)^2 \right ] \nonumber\\
+ \nabla \cdot \left [ m_s R^2 \int d^3w\, f_s \bv (\bv \cdot
\zun)^2 \right ] - m_s R^2 \sum_{s^\prime} \int d^3w \,
C_{ss^\prime} [ f_s, f_{s^\prime} ] (\bv \cdot \zun)^2,
\end{eqnarray}
where we have used $R\bB \times \zun = \nabla \psi$ and
$\nabla(R\zun) = (\nabla R) \zun - \zun (\nabla R)$. Taking a flux
surface average and the coarse grain averages over radius and time
in this equation, we find that the second and third terms in the
right side vanish to order $\epsilon_i^2 p_s R |\nabla \psi|$. The
last term only gives a neoclassical contribution, leaving the
turbulent transport of momentum
\begin{equation} \label{Pitb}
\left \langle \Pi_s^\mathrm{tb} \right \rangle_{\Delta t} = - c
\Bigg \langle \Bigg \langle \bigg \langle m_s R^2 \int d^3w\, \bar
h_s (\bw \cdot \zun + R\Omega_\zeta) \bigg ( \delta \bar \bE +
\frac{1}{c} \bw \times \delta \bar \bB \bigg ) \cdot \zun \bigg
\rangle_\psi \Bigg \rangle_{\Delta \psi} \Bigg \rangle_{\Delta t}
\sim \epsilon_i^2 p_s R|\nabla \psi|.
\end{equation}
\end{widetext}
Note that only the fluctuating piece of the distribution function
$\delta \bar f_s = \bar h_s - (Z_s e \bar \phi/T_s) f_{Ms}$
contributes to the turbulent transport of toroidal angular
momentum because the coarse grain averages satisfy $\langle
\langle \delta \bar f_s \rangle_{\Delta \psi} \rangle_{\Delta t} =
0$, $\langle \langle \delta \bar \bE \rangle_{\Delta \psi}
\rangle_{\Delta t} = 0$ and $\langle \langle \delta \bar \bB
\rangle_{\Delta \psi} \rangle_{\Delta t} = 0$. In addition, the
contribution from the Maxwell-Boltzmann response $- (Z_s e \bar
\phi/T_s) f_{Ms}$ vanishes upon integrating over velocity space
and over the spatial scales of the turbulence.

The turbulent contribution to the Maxwell stress is
\begin{equation} \label{PitbB}
\Pi_B^\mathrm{tb} = - \frac{1}{4\pi} \left \langle \left \langle R
(\delta \bar \bB \cdot \zun) (\delta \bar \bB \cdot \nabla \psi)
\right \rangle_\psi \right \rangle_{\Delta \psi}.
\end{equation}
This contribution is usually small for low $\beta$ plasmas, but it
can become important where the pressure gradient approaches the
ballooning threshold \cite{snyder01, jenko01}.

To study the different contributions to the turbulent transport of
momentum in \eq{Pitb} and \eq{PitbB}, we rewrite
$\Pi_s^\mathrm{tb}$ and $\Pi_B^\mathrm{tb}$ using $\delta \bar \bE
= - \nabla_\bot \bar \phi$, $\delta \bar \bB = \bar B_{||} \bun +
\nabla \bar A_{||} \times \bun$, $\bw \times \delta \bar \bB =
\nabla_\bot \bar A_{||} w_{||} - (\bw_\bot \cdot \nabla_\bot \bar
A_{||}) \bun + \bar B_{||} (\bw \times \bun)$ and $R\zun =
(I\bun/B) - (\bun \times \nabla \psi/B)$, giving
\begin{equation} \label{Pitb2}
\Pi_s^\mathrm{tb} = \pi_{s,||} + \pi_{s,\bot} + m_s \langle R^2
\rangle_\psi \Omega_\zeta g_s
\end{equation}
and
\begin{equation} \label{PitbB2}
\Pi_B^\mathrm{tb} = \pi_{B,||} + \pi_{B,\bot}.
\end{equation}
The transport of momentum has three contributions: a piece $\sum_s
m_s \langle R^2 \rangle_\psi \Omega_\zeta g_s$, where $g_s$ is
very similar to but not exactly the radial flux of particles, a
piece $\sum_s (\pi_{s,||} + \pi_{s,\bot})$ that is related to the
radial-toroidal component of the viscosity and the Reynolds
stress, and the Maxwell stress $\pi_{B,||} + \pi_{B, \bot}$. The
different contributions $\pi_{s, ||}$, $\pi_{s, \bot}$,
$\pi_{B,||}$ and $\pi_{B,\bot}$ are classified according to the
component of the momentum that they are transporting. The
components
\begin{eqnarray} \label{pipar}
\pi_{s,||} = \Bigg \langle \bigg \langle \frac{m_s I}{B} \int
d^3w\, \bar h_s \bigg \{ w_{||} \bar v_\mathrm{tb}^\psi +
\frac{1}{B} [ Iw_{||} \nonumber\\ - (\bw \times \bun) \cdot \nabla
\psi ] (\bw_\bot \cdot \nabla \bar A_{||}) \bigg \} \bigg
\rangle_\psi \Bigg \rangle_{\Delta \psi}
\end{eqnarray}
and
\begin{equation} \label{piBpar}
\pi_{B,||} = - \Bigg \langle \bigg \langle \frac{I}{4\pi B} \bar
B_{||} (\nabla \bar A_{||} \times \bun) \cdot \nabla \psi \bigg
\rangle_\psi \Bigg \rangle_{\Delta \psi}
\end{equation}
transport parallel momentum and dominate in a tokamak where the
magnetic field is mainly toroidal. Here the radial component of
the turbulent velocity is
\begin{equation}
\bar v_\mathrm{tb}^\psi = \left ( - \frac{c}{B} \nabla \bar \phi
\times \bun + \frac{w_{||}}{B} \nabla \bar A_{||} \times \bun -
\frac{\bar B_{||}}{B} \bw_\bot \right ) \cdot \nabla \psi.
\end{equation}
The components
\begin{equation} \label{piperp}
\pi_{s,\bot} = - \Bigg \langle \bigg \langle \frac{m_s}{B} \int
d^3w\, \bar h_s [ (\bw \times \bun) \cdot \nabla \psi ] \bar
v_\mathrm{tb}^\psi \bigg \rangle_\psi \Bigg \rangle_{\Delta \psi}
\end{equation}
and
\begin{equation} \label{piBperp}
\pi_{B,\bot} = - \Bigg \langle \bigg \langle \frac{1}{4\pi B}
(\nabla \bar A_{||} \cdot \nabla \psi) (\nabla \bar A_{||} \times
\bun) \cdot \nabla \psi \bigg \rangle_\psi \Bigg \rangle_{\Delta
\psi}
\end{equation}
transport perpendicular momentum and dominate for tokamaks where
the magnetic field is mainly poloidal. Finally, $g_s$ is
\begin{equation} \label{g}
g_s = \frac{1}{\langle R^2 \rangle_\psi} \Bigg \langle \bigg
\langle R^2 \int d^3w\, \bar h_s \bar v_\mathrm{tb}^\psi \bigg
\rangle_\psi \Bigg \rangle_{\Delta \psi}.
\end{equation}
In Appendix~\ref{app:flux} we evaluate $\pi_{s,||}$, $\pi_{s,
\bot}$, $g_s$, $\pi_{B, ||}$ and $\pi_{B, \bot}$ employing the
Fourier analyzed distribution function and electromagnetic field
obtained from the local formulation in section~\ref{sub:local}. In
what follows, the different contributions to $\pi_{s, ||}$,
$\pi_{s, \bot}$ and $g_s$ are classified according to which type
of fluctuation in the electromagnetic fields drives them. The flux
of parallel momentum due to species $s$ is $\pi_{s, ||} = \pi_{s,
||}^\phi + \pi_{s, ||}^{A_{||}} + \pi_{s, ||}^{B_{||}}$, with
\begin{eqnarray} \label{piparphikspace}
\pi_{s,||}^\phi = \frac{4 \pi^2 i m_s c I}{V^\prime} \sum_{
k_\psi, k_\alpha} k_\alpha \int d\theta\, \mathcal{J} \phi
(k_\psi, k_\alpha) \nonumber \\ \times \int dw_{||}\, d\mu\, h_s(-
k_\psi, - k_\alpha) w_{||} J_0(z_s),
\end{eqnarray}
\begin{eqnarray} \label{piparAparkspace}
\pi_{s,||}^{A_{||}} = - \frac{4 \pi^2 i m_s I}{V^\prime} \sum_{
k_\psi, k_\alpha} k_\alpha \int d\theta\, \mathcal{J} A_{||}
(k_\psi, k_\alpha) \nonumber \\ \times \int dw_{||}\, d\mu\, h_s
(-k_\psi, -k_\alpha) \Bigg [ w_{||}^2 J_0(z_s) - \mu B \frac{2 J_1
(z_s)}{z_s} \Bigg ]
\end{eqnarray}
and
\begin{eqnarray} \label{piparBparkspace}
\pi_{s,||}^{B_{||}} = \frac{4\pi^2 i m_s^2 c I}{Z_s e V^\prime}
\sum_{ k_\psi, k_\alpha} k_\alpha \int d\theta \mathcal{J} B_{||}
(k_\psi, k_\alpha) \nonumber \\ \times \int dw_{||}\, d\mu\, h_s
(-k_\psi, -k_\alpha) w_{||} \mu \frac{2 J_1 (z_s)}{z_s}.
\end{eqnarray}
Similarly $\pi_{s, \bot} = \pi_{s, \bot}^\phi + \pi_{s,
\bot}^{A_{||}} + \pi_{s, \bot}^{B_{||}}$, with
\begin{eqnarray} \label{piperpphikspace}
\pi_{s,\bot}^\phi = - \frac{4 \pi^2 m_s^2 c^2}{Z_s e V^\prime}
\sum_{ k_\psi, k_\alpha} k_\alpha \int d\theta\, \mathcal{J}
k^\psi \phi (k_\psi, k_\alpha) \nonumber\\\times \int dw_{||}\,
d\mu\, h_s (-k_\psi, -k_\alpha) \mu \frac{2J_1(z_s)}{z_s},
\end{eqnarray}
\begin{eqnarray} \label{piperpAparkspace}
\pi_{s,\bot}^{A_{||}} = \frac{4\pi^2 m_s^2 c}{Z_s e V^\prime}
\sum_{ k_\psi, k_\alpha} k_\alpha \int d\theta\, \mathcal{J}
k^\psi A_{||} (k_\psi, k_\alpha) \nonumber \\ \times \int
dw_{||}\, d\mu\, h_s (-k_\psi, -k_\alpha) w_{||} \mu
\frac{2J_1(z_s)}{z_s}
\end{eqnarray}
and
\begin{eqnarray} \label{piperpBparkspace}
\pi_{s,\bot}^{B_{||}} = - \frac{2\pi^2 m_s^3 c^2}{Z_s^2 e^2
V^\prime} \sum_{ k_\psi, k_\alpha} k_\alpha \int d\theta\,
\mathcal{J} k^\psi B_{||} (k_\psi, k_\alpha) \nonumber\\ \times
\int dw_{||}\, d\mu\, h_s (-k_\psi, -k_\alpha) \mu^2 G(z_s),
\end{eqnarray}
where $G(z_s) = [8J_1(z_s) + 4 z_s J_2(z_s) - 4 z_s J_0
(z_s)]/z_s^3$ and $k^\psi = \bk_\bot \cdot \nabla \psi = k_\psi
|\nabla \psi|^2 + k_\alpha \nabla \psi \cdot \nabla \alpha$ is the
projection of the wavevector on the radial direction. The
component $g_s = g_s^\phi + g_s^{A_{||}} + g_s^{B_{||}}$, where
\begin{eqnarray} \label{gphikspace}
g_s^\phi = \frac{4 \pi^2 i c}{V^\prime \langle R^2 \rangle_\psi}
\sum_{ k_\psi, k_\alpha} k_\alpha \int d\theta\, \mathcal{J} R^2 B
\phi (k_\psi, k_\alpha) \nonumber \\ \times \int dw_{||}\, d\mu\,
h_s(- k_\psi, - k_\alpha) J_0(z_s),
\end{eqnarray}
\begin{eqnarray} \label{gAparkspace}
g_s^{A_{||}} = - \frac{4 \pi^2 i}{V^\prime \langle R^2
\rangle_\psi} \sum_{k_\psi, k_\alpha} k_\alpha \int d\theta\,
\mathcal{J} R^2 B A_{||} (k_\psi, k_\alpha) \nonumber \\ \times
\int dw_{||}\, d\mu\, h_s(- k_\psi, - k_\alpha) w_{||} J_0(z_s)
\end{eqnarray}
and
\begin{eqnarray} \label{gBparkspace}
g_s^{B_{||}} = \frac{4 \pi^2 i m_s c}{Z_s e V^\prime \langle R^2
\rangle_\psi} \sum_{ k_\psi, k_\alpha} k_\alpha \int d\theta\,
\mathcal{J} R^2 B B_{||} (k_\psi, k_\alpha) \nonumber \\ \times
\int dw_{||}\, d\mu\, h_s(- k_\psi, - k_\alpha) \mu
\frac{2J_1(z_s)}{z_s}.
\end{eqnarray}
The Maxwell stress contribution to the transport of parallel
momentum is
\begin{equation} \label{piBparkspace}
\pi_{B, ||} = - \frac{i I}{2 V^\prime} \sum_{ k_\psi, k_\alpha}
k_\alpha \int d\theta\, \mathcal{J} B_{||} (k_\psi, k_\alpha)
A_{||} (-k_\psi, -k_\alpha).
\end{equation}
Finally, the Maxwell stress contribution to the transport of
perpendicular momentum is
\begin{equation} \label{piBperpkspace}
\pi_{B, \bot} = \frac{1}{2 V^\prime} \sum_{ k_\psi, k_\alpha}
k_\alpha \int d\theta\, \mathcal{J} k^\psi | A_{||} (k_\psi,
k_\alpha)|^2.
\end{equation}

\begin{figure}
\begin{center}
\includegraphics[width = 8 cm]{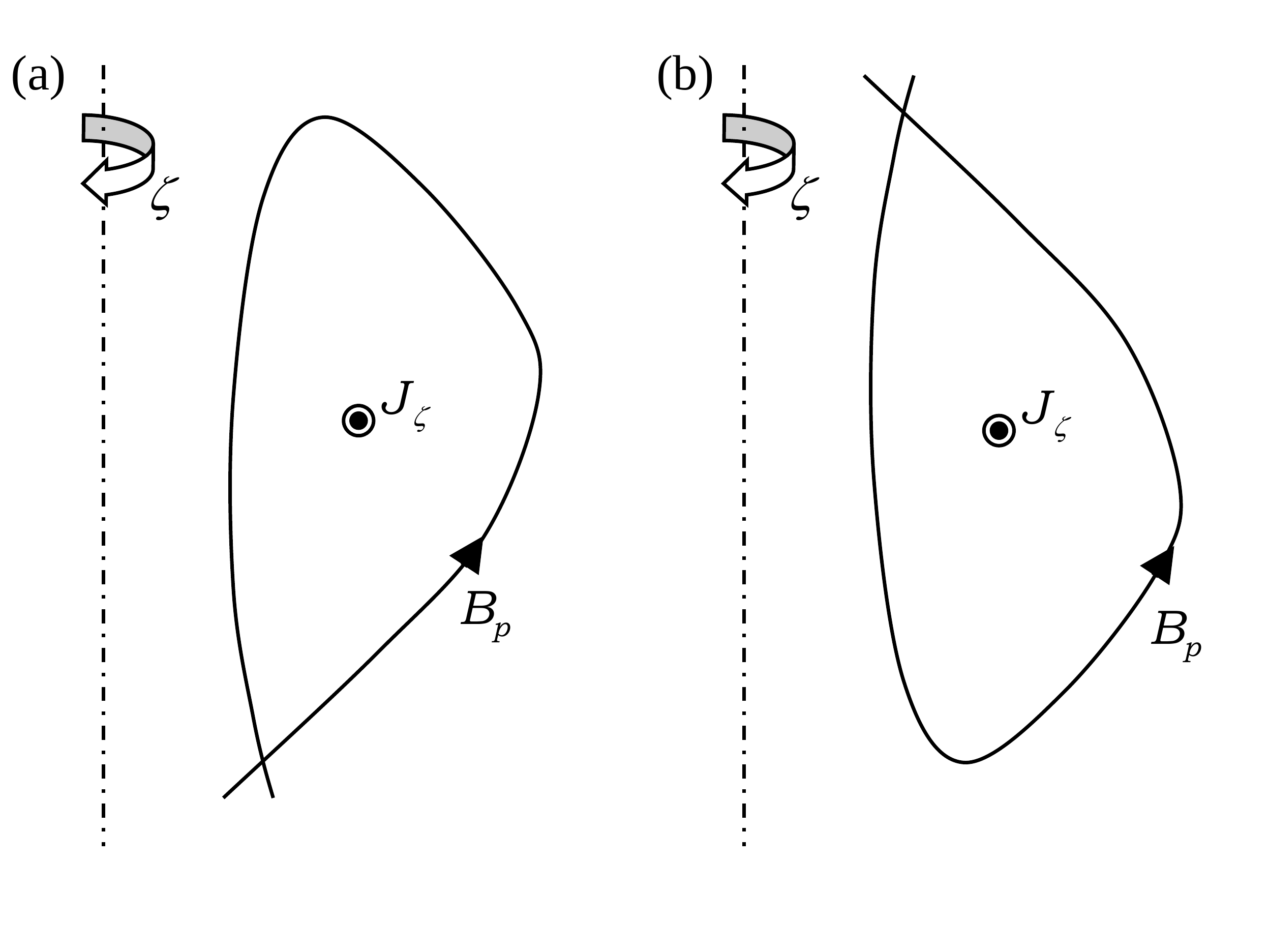}
\end{center}

\caption{Equilibrium (a) and its up-down reflection (b). The
dash-dot line is the axis of symmetry.} \label{updownreflection}
\end{figure}

\section{Symmetry of the local $\delta f$ gyrokinetic formulation} \label{sec:symmetry}
In this section we present two symmetries of the local $\delta f$
gyrokinetic formulation. Both of them involve an up-down
reflection as represented in Fig.~\ref{updownreflection}. Note
that the shape of the flux surface is an up-down reflection of the
original one, but the direction of the poloidal magnetic field is
unchanged. Thus, this reflection leaves the direction of the
toroidal plasma current unchanged. This up-down reflection does
not affect the toroidal magnetic field. In fact, one of the
symmetries that we present here will leave it unchanged whereas
the other reverses it. Importantly, for a given equilibrium, with
some chosen poloidal angle definition $\theta$, it is possible to
find a related poloidal angle $\theta$ for the up-down reflection
such that the functions of $\psi$ and $\theta$ given in
Table~\ref{table_geom} satisfy the relations given in that Table.
We prove this in Appendix~\ref{app:updown}.

\begin{table}
\caption{Equilibrium quantities under up-down reflection}
\label{table_geom}

\begin{tabular}{l l l}
\hline \hline Before up-down reflection & & After up-down reflection \\
\hline $|\nabla \psi|^2$, $\nabla \psi \cdot \nabla \theta$,
$|\nabla \theta|^2$, $B$ & & $|\nabla \psi|^2$, $\nabla \psi \cdot \nabla \theta$, $|\nabla \theta|^2$, $B$,\\
$\mathcal{J}$, $n_s$, $\Phi_0$, $p$, $\partial p/\partial \psi|_R$
& & $\mathcal{J}$, $n_s$, $\Phi_0$, $p$, $\partial
p/\partial \psi|_R$ \vspace{1.5mm}\\
$\bB \cdot \nabla \theta$ & & $- \bB \cdot \nabla \theta$ \\
\hline \hline
\end{tabular}
\end{table}

There are two transformations, which we will call transformations
I and II, that leave the local $\delta f$ gyrokinetic system of
equations unchanged. They are given in Table~\ref{table_change}.
To prove that these symmetries hold, it is necessary to know how
certain coefficients in the gyrokinetic equations transform under
up-down reflection. This information is contained in
Tables~\ref{table_geom} and \ref{table_geom_2}.
Table~\ref{table_geom_2}, derived in Appendix~\ref{app:updown2},
is for the coefficients that contain $\nabla \alpha$. Since they
depend on $I$ and $\partial \Omega_\zeta/\partial \psi$, they
change differently under transformation I or II. With the
information in Tables~\ref{table_geom} and \ref{table_geom_2}, it
is tedious but straightforward to show that the formulation given
in subsection~\ref{sub:local} is invariant under the
transformations in Table~\ref{table_change}. The symmetry under
transformation I is the one that is closer to the the symmetry
discovered in Ref.~\cite{peeters05}. Since in
Ref.~\cite{peeters05} some of the effects that we are considering
here were not treated, the transformation is not exactly the same.
In particular, Ref.~\cite{peeters05} neglected $A_{||}$ and
$B_{||}$ and proposed a transformation composed only of the
up-down reflection and the change $w_{||} \rightarrow - w_{||}$.
This symmetry is only valid for the linearized equations and for
the particular case $k_\psi = 0$, which is the most commonly
studied in local $\delta f$ gyrokinetic linear theory. In
addition, Ref.~\cite{peeters05} did not state that $h_s$ and
$\phi$ should change sign under the transformation. The reason is
that the symmetry of the linearized equations cannot determine if
$h_s$, $\phi$, $A_{||}$ or $B_{||}$ must change sign. Only the
nonlinear term can determine that.

\begin{table}
\caption{Transformation I and II that leave the local $\delta f$
gyrokinetic system of equations unchanged.} \label{table_change}

\begin{tabular}{ c l l l l}
\hline \hline & & Before transformation & & After transformation\\
\hline \multirow{4}{*}{I} & & Geometry & & Up-down reflected geometry \vspace{1.5mm} \\
 & & $\mu$, $k_\alpha$, $A_{||}$ & & $\mu$, $k_\alpha$, $A_{||}$ \vspace{1.5mm} \\
 & & $\Omega_\zeta$, $\partial \Omega_\zeta/\partial \psi$, $w_{||}$, $k_\psi$, & & $- \Omega_\zeta$, $- \partial \Omega_\zeta/\partial \psi$, $- w_{||}$, $- k_\psi$, \\
 & & $h_s$, $\phi$, $B_{||}$ & & $- h_s$, $- \phi$, $- B_{||}$ \\
\hline \multirow{3}{*}{II} & & Geometry & & Up-down reflected geometry \vspace{1.5mm} \\
 & & $\mu$, $k_\psi$, $k_\alpha$, $h_s$, $\phi$, $B_{||}$ & & $\mu$, $k_\psi$, $k_\alpha$, $h_s$, $\phi$, $B_{||}$ \vspace{1.5mm} \\
 & & $I$, $w_{||}$, $A_{||}$ & & $- I$, $- w_{||}$, $- A_{||}$ \\
\hline \hline
\end{tabular}
\end{table}

\section{Consequences for the transport of toroidal angular
momentum} \label{sec:conclusion}

According to transformation I in Table~\ref{table_change}, if $h_s
(k_\psi, k_\alpha, \theta, w_{||}, \mu, t)$, $\phi(k_\psi,
k_\alpha, \theta, t)$, $A_{||} (k_\psi, k_\alpha, \theta, t)$ and
$B_{||} (k_\psi, k_\alpha, \theta, t)$ are solutions of the
equations in subsection~\ref{sub:local} in a given equilibrium,
then $- h_s (- k_\psi, k_\alpha, \theta, - w_{||}, \mu, t)$, $-
\phi(- k_\psi, k_\alpha, \theta, t)$, $A_{||} (- k_\psi, k_\alpha,
\theta, t)$ and $- B_{||} (- k_\psi, k_\alpha, \theta, t)$ are
solutions in the up-down reflection with $\Omega_\zeta \rightarrow
- \Omega_\zeta$ and $\partial \Omega_\zeta/\partial \psi
\rightarrow - \partial \Omega_\zeta/\partial \psi$. This is
important for transport of momentum because the radial transport
of toroidal angular momentum, given in
subsection~\ref{sub:fluxes}, changes sign under transformation I.
The contributions $\pi_{s, ||}^\phi$, $\pi_{s, ||}^{A_{||}}$,
$\pi_{s, ||}^{B_{||}}$, $\pi_{s, \bot}^\phi$, $\pi_{s,
\bot}^{A_{||}}$, $\pi_{s, \bot}^{B_{||}}$, $\pi_{B, ||}$ and
$\pi_{B,\bot}$ change sign. The contributions $g_s^\phi$,
$g_s^{A_{||}}$ and $g_s^{B_{||}}$ remain the same, but since they
appear in the momentum flux multiplied by $\Omega_\zeta$ their
total contribution also changes sign. Thus, for any initial
condition $h_s (k_\psi, k_\alpha, \theta, w_{||}, \mu, t = 0)$,
there is another initial condition $- h_s (- k_\psi, k_\alpha,
\theta, - w_{||}, \mu, t = 0)$ that gives the opposite $\pi_{s,
||}$, $\pi_{s, \bot}$, $\pi_{B,||}$ and $\pi_{B, \bot}$ and the
same $g_s$ for the up-down reflection with $\Omega_\zeta
\rightarrow -\Omega_\zeta$ and $\partial \Omega_\zeta/\partial
\psi \rightarrow - \partial \Omega_\zeta/\partial \psi$. Since
$\pi_{s, ||}$, $\pi_{s, \bot}$, $g_s$, $\pi_{B, ||}$ and
$\pi_{B,\bot}$ are time averaged over times longer than the
decorrelation time, they are independent of the initial condition,
and the radial transport of momentum must change sign under
up-down reflection and the change $\Omega_\zeta \rightarrow
-\Omega_\zeta$ and $\partial \Omega_\zeta/\partial \psi
\rightarrow - \partial \Omega_\zeta/\partial \psi$. Because the
symmetry changes the sign of each contribution $\pi_{s, ||}^\phi$,
$\pi_{s, ||}^{A_{||}}$, $\pi_{s, ||}^{B_{||}}$, $\pi_{s,
\bot}^\phi$, $\pi_{s, \bot}^{A_{||}}$, $\pi_{s, \bot}^{B_{||}}$,
$\pi_{B, ||}$ and $\pi_{B, \bot}$ independently, this property
holds for each one of them independently. The same argument is
valid for $g_s^\phi$, $g_s^{A_{||}}$ and $g_s^{B_{||}}$. In the
particular case of up-down symmetric tokamaks, this symmetry
implies that the sign of the radial transport of momentum changes
when the sign of $\Omega_\zeta$ and $\partial
\Omega_\zeta/\partial \psi$ is changed, and when $\Omega_\zeta =
0$ and $\partial \Omega_\zeta/\partial \psi = 0$, it must vanish.
A consequence of this result is that reversing the momentum input
in up-down symmetric tokamaks exactly reverses the rotation
profile $\Omega_\zeta (\psi)$.

\begin{table}
\caption{Changes under transformations I and II.}
\label{table_geom_2}

\begin{tabular}{ c l l l l}
\hline \hline & & Before transformation & & After transformation \\
\hline \multirow{3}{*}{I} & & $|\nabla \alpha|^2$, $(\nabla \alpha \times \bB) \cdot \nabla \theta$, & & $|\nabla \alpha|^2$, $(\nabla \alpha \times \bB) \cdot \nabla \theta$, \\
 & & $(\nabla R \times \zun) \cdot \nabla \alpha$ & & $(\nabla R \times \zun) \cdot \nabla \alpha$ \vspace{1.5mm} \\
 & & $\nabla \psi \cdot \nabla \alpha$, $\nabla \theta \cdot \nabla \alpha$ & & $- \nabla \psi \cdot \nabla \alpha$, $- \nabla \theta \cdot \nabla \alpha$ \\
\hline \multirow{3}{*}{II} & & $\nabla \psi \cdot \nabla \alpha$, $\nabla \theta \cdot \nabla \alpha$, $|\nabla \alpha|^2$, & & $\nabla \psi \cdot \nabla \alpha$, $\nabla \theta \cdot \nabla \alpha$, $|\nabla \alpha|^2$, \\
 & & $(\nabla \alpha \times \bB) \cdot \nabla \theta$ & & $(\nabla \alpha \times \bB) \cdot \nabla \theta$ \vspace{1.5mm} \\
 & & $(\nabla R \times \zun) \cdot \nabla \alpha$ & & $- (\nabla R \times \zun) \cdot \nabla \alpha$\\
\hline \hline
\end{tabular}
\end{table}

For up-down asymmetric configurations, there is turbulent
transport of momentum even for $\Omega_\zeta = 0$ and $\partial
\Omega_\zeta/\partial \psi = 0$ \cite{camenen09}. This momentum
flux is strongly dependent on the temperature and density
gradients that drive the turbulence. An explicit expression for
the turbulent momentum flux due to up-down asymmetry has not been
obtained, but it is possible to comment on the size of this
contribution for small up-down asymmetry. We distinguish two types
of equilibrium quantities that enter in the coefficients of the
gyrokinetic system of equations in subsection~\ref{sub:local}: the
quantities $Q(\psi, \theta)$ for which, in an up-down symmetric
configuration, there exists a poloidal angle definition $\theta$
such that $Q(\psi, \theta) = Q(\psi, -\theta)$, and the quantities
$\overline{Q} (\psi, \theta)$, for which $\overline{Q}(\psi,
\theta) = - \overline{Q}(\psi, -\theta)$. The former quantities
are $Q (\psi, \theta) = \{ |\nabla \psi|^2, |\nabla \theta|^2, B,
\bB \cdot \nabla \theta, \mathcal{J}, n_s, \Phi_0, p,\partial
p/\partial \psi|_R \}$ and the latter quantities are $\overline{Q}
(\psi, \theta) = \{ \nabla \psi\cdot \nabla \theta \}$. Any
equilibrium quantity can be divided in its up-down symmetric and
antisymmetric pieces, given by $Q^S (\psi, \theta) = [Q (\psi,
\theta) + Q(\psi, - \theta)]/2$ and $Q^A (\psi, \theta) = [Q
(\psi, \theta) - Q(\psi, - \theta)]/2$ for $Q$, and by
$\overline{Q}^S (\psi, \theta) = [\overline{Q} (\psi, \theta) +
\overline{Q}(\psi, - \theta)]/2$ and $\overline{Q}^A (\psi,
\theta) = [\overline{Q} (\psi, \theta) - \overline{Q}(\psi, -
\theta)]/2$ for $\overline{Q}$. In a tokamak close to up-down
symmetry, there is a poloidal angle definition $\theta$ for which
$Q^A \ll Q^S$ and $\overline{Q}^S \ll \overline{Q}^A$. In the
gyrokinetic equation \eq{kinetickspace} and the Maxwell equations
\eq{qnkspace}-\eq{perppresskspace}, the quantities $Q$ and
$\overline{Q}$ enter in different coefficients. For small up-down
symmetry, we can expand in $Q^A/Q^S \ll 1$ and $\overline{Q}^S/
\overline{Q}^A \ll 1$, leaving to lowest order a system of
gyrokinetic equations that only depend on $Q^S$ and
$\overline{Q}^A$. In this case, the lowest order solutions
$h_s^{(0)}$, $\phi^{(0)}$, $A_{||}^{(0)}$ and $B_{||}^{(0)}$ have
the symmetry properties of the up-down symmetric case and the
momentum flux is zero. For example,
\begin{eqnarray}
\pi_{s,||}^{\phi(0)} = \frac{4 \pi^2 i m_s c I}{V^\prime} \sum_{
k_\psi, k_\alpha} k_\alpha \int d\theta\, \mathcal{J}^S \phi^{(0)}
(k_\psi, k_\alpha) \nonumber \\ \times \int dw_{||}\, d\mu\,
h_s^{(0)} (- k_\psi, - k_\alpha) w_{||} J_0(z_s^S) = 0,
\end{eqnarray}
where by using the superscript $S$ we have emphasized that to this
order $\mathcal{J}$ and $z_s$ are evaluated using only the up-down
symmetric coefficients $Q^S$ and $\overline{Q}^A$. The next order
correction to the system of gyrokinetic equations is linear in the
next order corrections $h_s^{(1)}$, $\phi^{(1)}$, $A_{||}^{(1)}$
and $B_{||}^{(1)}$ to the lowest order solutions, and contains the
antisymmetric coefficients $Q^A$ and $\overline{Q}^S$. As a
result,
\begin{equation}
\frac{h_s^{(1)}}{h_s^{(0)}} \sim \frac{\phi^{(1)}}{\phi^{(0)}}
\sim \frac{A_{||}^{(1)}}{A_{||}^{(0)}} \sim
\frac{B_{||}^{(1)}}{B_{||}^{(0)}} \sim \frac{Q^A}{Q^S} \sim
\frac{\overline{Q}^S}{\overline{Q}^A} \ll 1.
\end{equation}
Since the lowest order momentum flux is zero, the contribution due
to these higher order correction becomes the only contribution.
For example, to next order we find
\begin{eqnarray}
\pi_{s,||}^{\phi(1)} = \frac{4 \pi^2 i m_s c I}{V^\prime} \sum_{
k_\psi, k_\alpha} k_\alpha \int d\theta\, \mathcal{J}^S \phi^{(1)}
(k_\psi, k_\alpha) \nonumber \\ \times \int dw_{||}\, d\mu\,
h_s^{(0)} (- k_\psi, - k_\alpha) w_{||} J_0(z_s^S) \nonumber\\ +
\frac{4 \pi^2 i m_s c I}{V^\prime} \sum_{ k_\psi, k_\alpha}
k_\alpha \int
d\theta\, \mathcal{J}^S \phi^{(0)} (k_\psi, k_\alpha) \nonumber \\
\times \int dw_{||}\, d\mu\, h_s^{(1)} (- k_\psi, - k_\alpha)
w_{||} J_0(z_s^S) \nonumber\\ + \frac{4 \pi^2 i m_s c I}{V^\prime}
\sum_{ k_\psi, k_\alpha} k_\alpha \int d\theta\, \mathcal{J}^A
\phi^{(0)} (k_\psi, k_\alpha) \nonumber \\ \times \int dw_{||}\,
d\mu\, h_s^{(0)} (- k_\psi, - k_\alpha) w_{||} J_0(z_s^S)
\nonumber\\ - \frac{4 \pi^2 i m_s c I}{V^\prime} \sum_{ k_\psi,
k_\alpha} k_\alpha \int
d\theta\, \mathcal{J}^S \phi^{(0)} (k_\psi, k_\alpha) \nonumber \\
\times \int dw_{||}\, d\mu\, h_s^{(0)} (- k_\psi, - k_\alpha)
w_{||} J_1 (z_s^S) z_s^A,
\end{eqnarray}
where the superscript $A$ in $\mathcal{J}^A$ and $z_s^A$ indicates
that these are the next order corrections in $Q^A/Q^S \ll 1$ and
$\overline{Q}^S/\overline{Q}^A \ll 1$ to the coefficients
$\mathcal{J}$ and $z_s$, and as a consequence, they contain $Q^A$
and $\overline{Q}^S$. Note that $\pi_{s,||}^{\phi(1)}$ contains
terms that have exactly the opposite symmetry to the symmetry that
made the lowest order contribution $\pi_{s,||}^{\phi(0)}$ zero.
Thus, the contribution to momentum flux due to up-down asymmetry
is proportional to the asymmetric component of the equilibrium
coefficients, that is, proportional to $Q^A$ and $\overline{Q}^S$.

\begin{figure*}
\begin{center}
\includegraphics[width = 8.5cm]{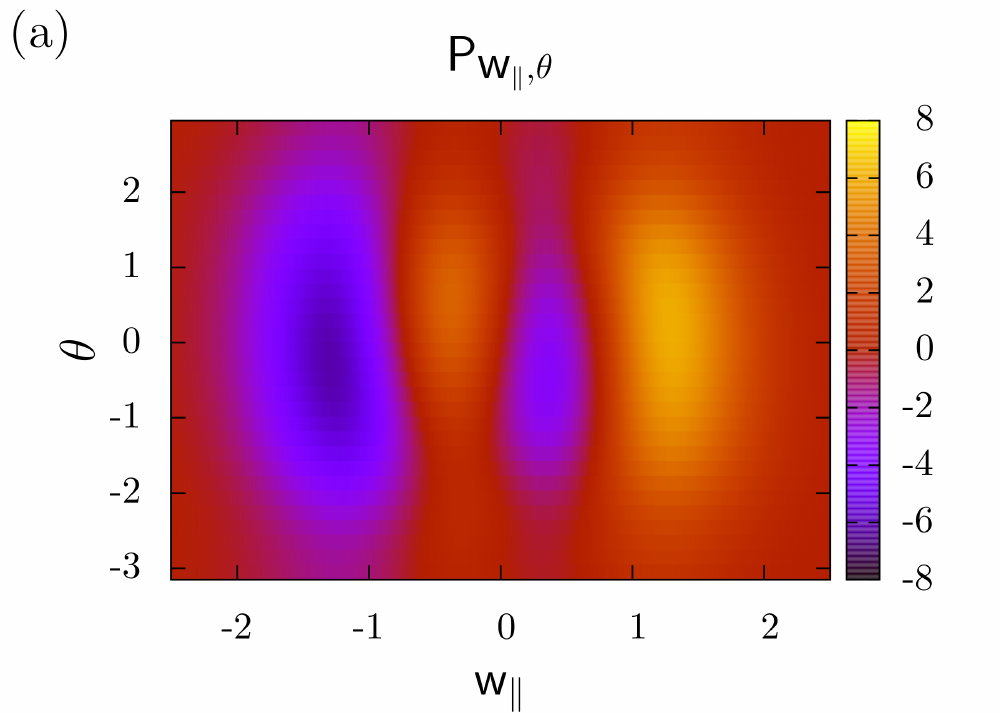}
\includegraphics[width = 8.5cm]{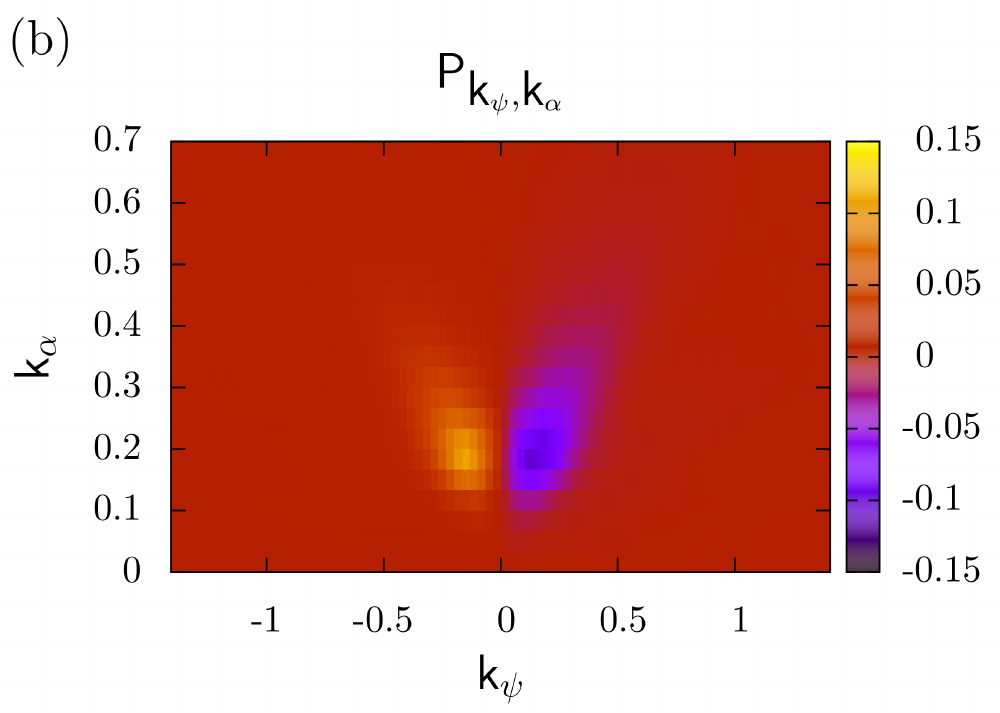}
\includegraphics[width = 8.5cm]{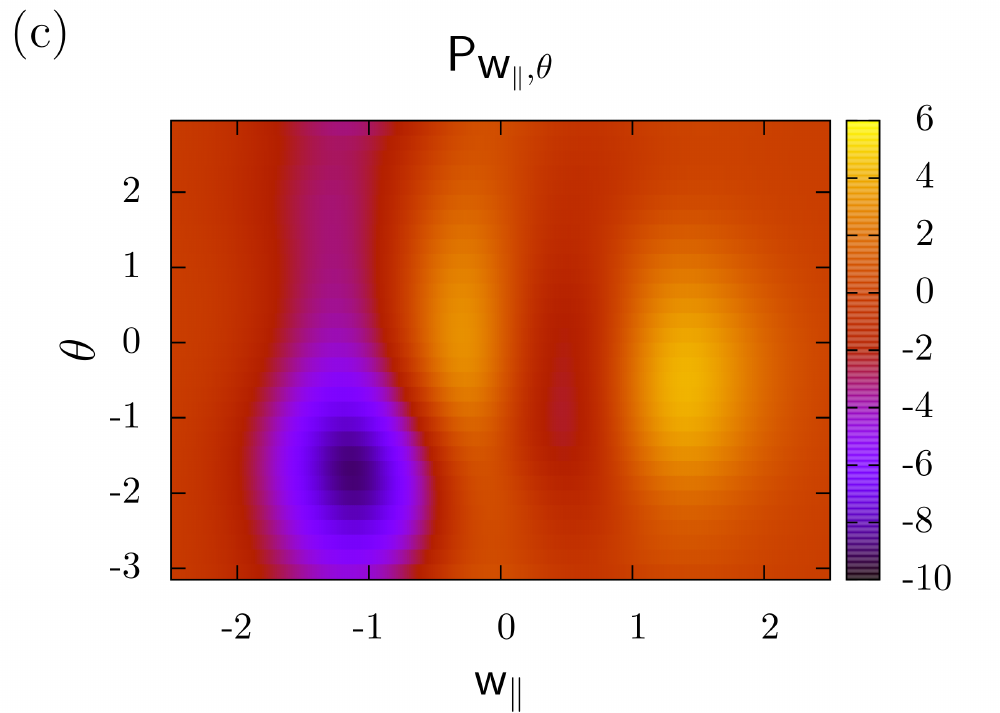}
\includegraphics[width = 8.5cm]{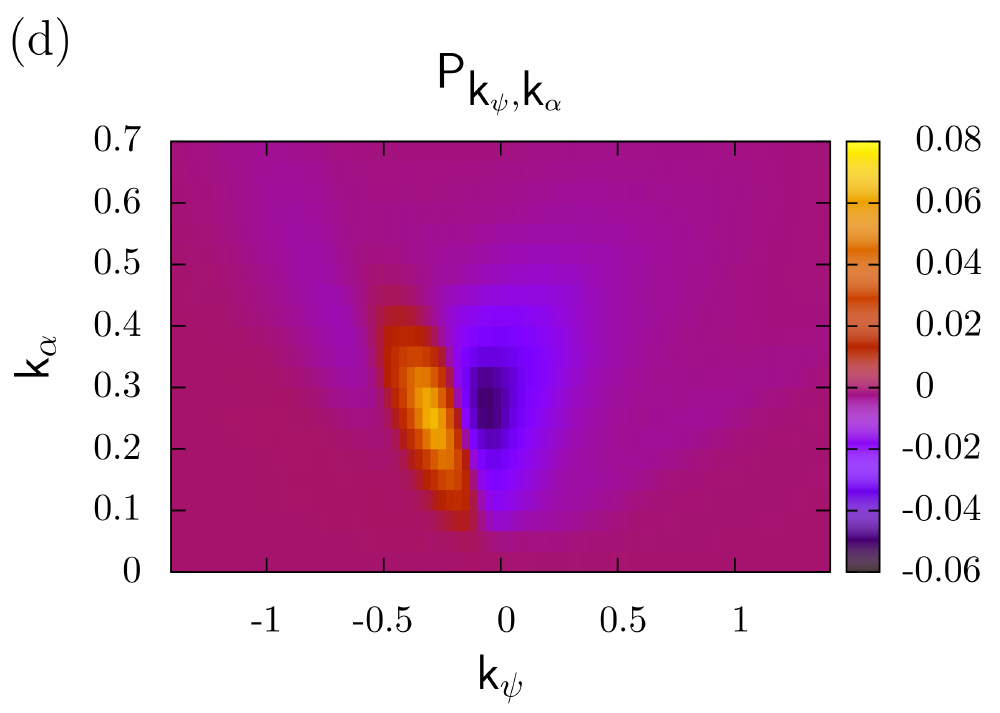}
\includegraphics[width = 8.5cm]{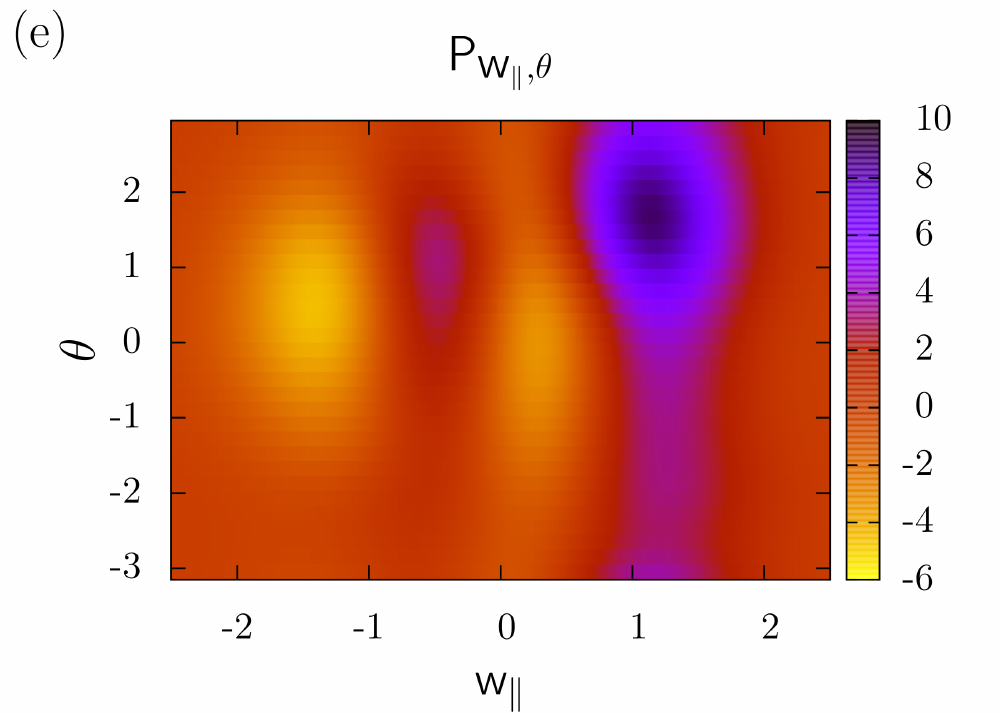}
\includegraphics[width = 8.5cm]{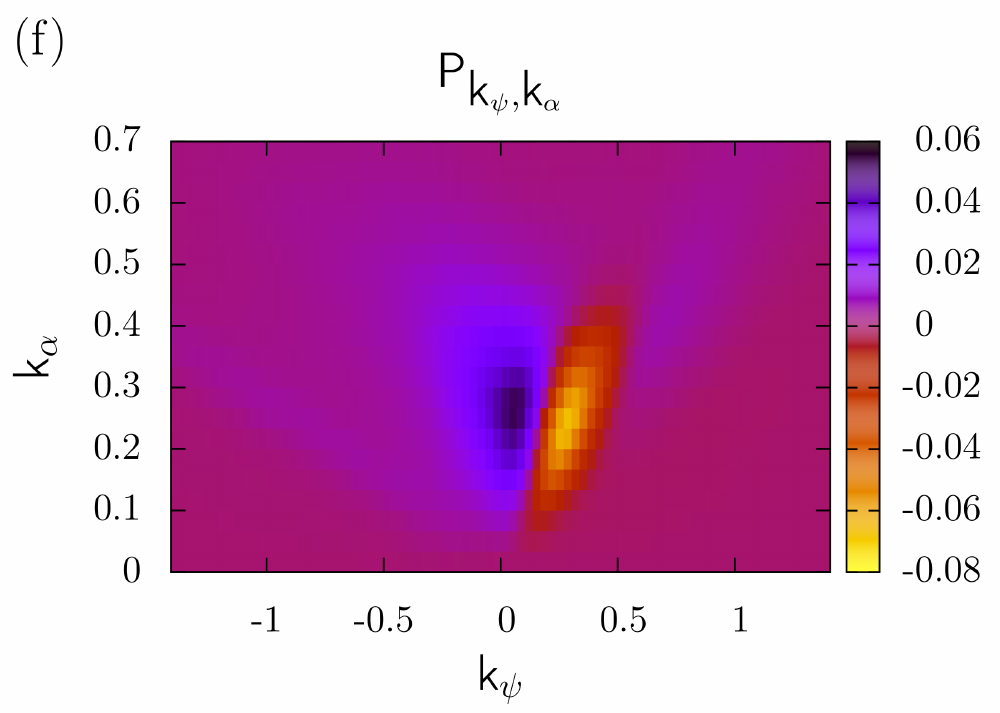}
\end{center}

\caption{(Color online) Contributions to momentum transport for
different values of $\partial \Omega_\zeta/\partial \psi$: (a)-(b)
$\partial \Omega_\zeta/\partial \psi = 0$; (c)-(d) $\partial
\Omega_\zeta/\partial \psi = 0.2\, q^2 v_{ti}/\varepsilon^2 R_0^3
B_0$; and (e)-(f) $\partial \Omega_\zeta/\partial \psi = - 0.2\,
q^2 v_{ti}/\varepsilon^2 R_0^3 B_0$. Note that in figures (e) and
(f) the color map has been inverted to aid the comparison with
figures (c) and (d).} \label{fig:piplots}
\end{figure*}

To demonstrate that the symmetry based on transformation I is
observed in local $\delta f$ gyrokinetic simulations, the local
$\delta f$ gyrokinetic code \texttt{GS2}~\cite{dorland00} was used
to solve the gyrokinetic system of equations \eq{kinetickspace}
and \eq{qnkspace} for an up-down symmetric equilibrium with
$\Omega_{\zeta} = 0$, but with three different values of $\partial
\Omega_{\zeta}/ \partial \psi$: zero and $\pm 0.2\, q^2
v_{ti}/\varepsilon^2 R_0^3 B_0$. Here $v_{ti} = \sqrt{T_i/m_i}$,
$T_i$ and $m_i$ are the ion thermal speed, temperature and mass,
$R_0$ and $B_0$ are the major radius and the magnetic field at the
magnetic axis, $q = 1.4$ is the safety factor, and $\varepsilon =
0.18$ is the inverse aspect ratio. All other equilibrium
parameters correspond to the Cyclone base case~\cite{dimits00}.
Only electrostatic fluctuations are considered and a modified
Boltzmann response \cite{hammett93} is employed for the electrons.
From the solutions for ${h}_i$, we calculate the dimensionless
function $\pivar (\kvar_\psi, \kvar_\alpha, \theta, \wvar_{||})$,
defined as
\begin{eqnarray}
\pivar = \frac{2\pi c k_{\alpha}
\phi(k_{\psi},k_{\alpha})}{\epsilon_i^2 R_0 n_i v_{ti} \left
\langle \left|\nabla\psi\right|\right \rangle_{\psi}} \int d\mu\,
h_i (-k_{\psi},-k_{\alpha}) \nonumber \\ \times \left(i I w_{||}
J_0(z_i) - \frac{m_i c}{e}k^{\psi}\mu
\frac{2J_1(z_i)}{z_i}\right),
\end{eqnarray}
where $n_i$, $p_i = n_i T_i$ and $\rho_i = m_i c v_{ti}/e B$ are
ion density, pressure and thermal gyroradius, and $\epsilon_i =
\rho_i/R_0$. The function $\pivar$ can be understood as the
dimensionless contribution to the radial momentum flux due to
fluctuations at the poloidal angle $\theta$ with dimensionless
wavelength $\kvar_\psi = k_\psi \epsilon_i (\varepsilon R_0^2
B_0/q)$ and $\kvar_\alpha = k_\alpha \epsilon_i (q/\varepsilon)$
when interacting with particles with dimensionless parallel
velocity $\wvar_{||} = w_{||}/v_{ti}$. The function $\pivar$ is
related to the turbulent toroidal angular momentum flux by
\begin{eqnarray}
\frac{\langle \pi^\phi_{i, ||} + \pi^\phi_{i, \bot}
\rangle_{\Delta t}}{\epsilon_i^2 R_0 p_i \left \langle \left
|\nabla\psi \right| \right \rangle_{\psi}} = \frac{2\pi}{V^\prime}
\sum_{\kvar_{\psi}, \kvar_{\alpha}} \int d\theta \mathcal{J} \int
d\wvar_{||} \pivar.
\end{eqnarray}
From the symmetry under transformation I in
Table~\ref{table_change}, we find that $\langle
\pivar(\kvar_{\psi}, \kvar_{\alpha}, \wvar_{||}, \theta)
\rangle_{\Delta t} = - \langle \pivar(-\kvar_{\psi}, \kvar_\alpha,
-\wvar_{||}, -\theta) \rangle_{\Delta t}$ for the case with
$\partial \Omega_{\zeta} / \partial\psi = 0$. The change in sign
in the angle $\theta$ is due to the up-down reflection. As proven
in Appendix~\ref{app:updown}, for every angle $\theta$ defined for
an equilibrium, there is an angle $\theta$ for the up-down
reflection such that Table~\ref{table_geom} is satisfied. In this
case, for which the up-down reflection is the flux surface itself,
we choose the poloidal angle definition $\theta$ so that the
corresponding poloidal angle for the up-down reflection is simply
$- \theta$. The result $\langle \pivar(\kvar_{\psi},
\kvar_{\alpha}, \wvar_{||}, \theta) \rangle_{\Delta t} = - \langle
\pivar(-\kvar_{\psi}, \kvar_\alpha, -\wvar_{||}, -\theta)
\rangle_{\Delta t}$ is confirmed by Figs.~\ref{fig:piplots}(a)
and~\ref{fig:piplots}(b), which clearly show that the integrals of
the function $\pivar (\kvar_\psi, \kvar_\alpha, \theta,
\wvar_{||})$
\begin{equation}
\pivar_{\wvar_{||}, \theta} (\wvar_{||},\theta)= \left \langle
\sum_{\kvar_{\psi}, \kvar_{\alpha}} \pivar \right \rangle_{\Delta
t}
\end{equation}
and
\begin{equation}
\pivar_{\kvar_\psi, \kvar_\alpha} (\kvar_{\psi}, \kvar_{\alpha}) =
\left \langle \frac{2\pi}{V^\prime} \int d\theta \mathcal{J} \int
d\wvar_{||} \pivar \right \rangle_{\Delta t}
\end{equation}
satisfy the relations
\begin{equation}
\pivar_{\wvar_{||}, \theta} (\wvar_{||},\theta) = -
\pivar_{\wvar_{||}, \theta} (- \wvar_{\parallel},-\theta)
\end{equation}
and
\begin{equation}
\pivar_{\kvar_\psi, \kvar_\alpha} (\kvar_{\psi}, \kvar_{\alpha}) =
- \pivar_{\kvar_\psi, \kvar_\alpha} (-\kvar_{\psi},
\kvar_{\alpha}).
\end{equation}
This symmetry is broken for the cases with finite
$\partial\Omega_{\zeta}/\partial\psi$, which are shown in
Figs.~\ref{fig:piplots}(c)-\ref{fig:piplots}(f). Observe that the
symmetry also gives that the case with negative $\partial
\Omega_\zeta/\partial \psi$ has $\textsf{P}_{-} (\kvar_\psi,
\kvar_\alpha, \wvar_{||}, \theta) = - \textsf{P}_{+} (-
\kvar_\psi, \kvar_\alpha, - \wvar_{||}, - \theta)$, where
$\textsf{P}_{+} (\kvar_\psi, \kvar_\alpha, \wvar_{||}, \theta)$ is
the function $\textsf{P}$ for the case with positive $\partial
\Omega_\zeta/\partial \psi$, and $\textsf{P}_{-} (\kvar_\psi,
\kvar_\alpha, \wvar_{||}, \theta)$ is the function $\textsf{P}$
for the negative $\partial \Omega_\zeta/\partial \psi$. This is
confirmed by comparing Fig.~\ref{fig:piplots}(c) with
\ref{fig:piplots}(e), and Fig.~\ref{fig:piplots}(d) with
\ref{fig:piplots}(f). As a consequence of these symmetries, the
case with $\partial\Omega_{\zeta}/\partial\psi=0$ has $\langle
\pi_{i,||}^\phi + \pi_{i, \bot}^\phi \rangle_{\Delta t} = 0$,
while the other cases have $\langle \pi_{i,||}^\phi + \pi_{i,
\bot}^\phi \rangle_{\Delta t} = \pm 2.6 \, \epsilon_i^2 R_0 p_i
\langle |\nabla \psi| \rangle_\psi$.

Following the same reasoning as for transformation I,
transformation II in Table~\ref{table_change} implies that the
transport of momentum remains the same for an up-down reflection
and a reversal of the toroidal magnetic field $B_\zeta = I/R$.
This means then that for an up-down symmetric tokamak, the
transport of momentum cannot depend on the direction of the
toroidal magnetic field. For example, the Coriolis pinch
\cite{peeters07} will remain the same under a reversal of the
toroidal magnetic field.

These symmetries of the lowest order gyrokinetic formulation have
very important consequences for the modeling of spontaneous
rotation profiles in tokamaks without net momentum input. When no
momentum is injected, the radial momentum flux across any flux
surface within the separatrix must vanish, i.e., $\Pi (\psi) = 0$.
In the absence of up-down asymmetry, the radial transport of
momentum given by the lowest order gyrokinetic model vanishes for
$\Omega_\zeta = 0$ and $\partial \Omega_\zeta/\partial \psi = 0$.
Thus, for zero or negligible velocity in the edge, the solution to
the equation $\Pi (\Omega_\zeta, \partial \Omega_\zeta/\partial
\psi) = 0$ is $\Omega_\zeta (\psi) = 0$ everywhere. To obtain
$\Omega_\zeta (\psi) \neq 0$ with the lowest order gyrokinetic
formulation, we need to assume that there is some up-down
asymmetry \cite{camenen09}. This limits the applicability of the
lowest order formulation to studies of spontaneous rotation in
very up-down asymmetric configurations. When higher order terms
are considered \cite{parra10a, parra11a}, more mechanisms for the
generation of spontaneous rotation arise.

\acknowledgments The authors are grateful for the hospitality of
the Isaac Newton Institute for Mathematical Sciences at University
of Cambridge. This article is the final result of a series of
conversations at the summer programme ``Gyrokinetics in Laboratory
and Astrophysical Plasmas" hosted there.

This work was supported in part by the postdoctoral fellowship
programme of UK EPSRC, the JRF programme of Christ Church at
University of Oxford, the Oxford-Culham Fusion Research Fellowship
and the Leverhulme Trust Network for Magnetized Plasma Turbulence.

\appendix

\section{Collision operator} \label{app:collision}

The linearized collision operator is given by
\begin{equation}
\bar C_{ss^\prime}^{(\ell)} = \nabla_w \cdot
(\bar{\Gammabf}_{ss^\prime}^\lambda +
\bar{\Gammabf}_{ss^\prime}^\varepsilon +
\bar{\Gammabf}_{ss^\prime}^R).
\end{equation}
The different contributions to the collision operator are the
pitch-angle scattering
\begin{equation}
\bar{\Gammabf}_{ss^\prime}^\lambda = \nu^\lambda_{ss^\prime} (w^2
\matI - \bw \bw) \cdot \nabla_w \bar h_s,
\end{equation}
the energy scattering
\begin{equation}
\bar{\Gammabf}_{ss^\prime}^\varepsilon =
\nu^\varepsilon_{ss^\prime} \, \bw \left ( \bw \cdot \nabla_w \bar
h_s + \frac{m_s w^2}{T_{s^\prime}} \bar h_s \right ),
\end{equation}
and the integral contribution
\begin{equation}
\bar{\Gammabf}_{ss^\prime}^R = - \frac{m_s \gamma_{ss^\prime}
f_{Ms}}{T_s} \left ( \bw \cdot \nabla_w \nabla_w \bar H_{s^\prime}
+ \frac{T_s}{m_{s^\prime}} \nabla_w \bar L_{s^\prime} \right ),
\end{equation}
where the Rosenbluth potentials $\bar H_{s^\prime}$ and $\bar
L_{s^\prime}$ are obtained by solving the differential equations
$\nabla_w^2 \bar L_{s^\prime} = - 8\pi \bar h_{s^\prime}$ and
 $\nabla_w^2 \bar H_{s^\prime} = \bar L_{s^\prime}$.
Here $\gamma_{ss^\prime} = 2\pi Z^2_s Z^2_{s^\prime} e^4 \ln
\Lambda/m_s^2$, $\ln \Lambda$ is Coulomb's logarithm,
\begin{equation}
\nu^\lambda_{ss^\prime}(w) = \frac{\gamma_{ss^\prime} n_{s^\prime}
m_{s^\prime}^{3/2}}{2^{3/2} T_{s^\prime}^{3/2}} \Bigg ( \frac{2
\overline{w}_{s^\prime}^2 -1}{2 \overline{w}_{s^\prime}^5} \Phi(
\overline{w}_{s^\prime} ) + \frac{1}{2\overline{w}_{s^\prime}^4}
\frac{d\Phi}{d\overline{w}_{s^\prime}} \Bigg )
\end{equation}
is the pitch-angle scattering collision frequency, and
\begin{equation}
\nu^\varepsilon_{ss^\prime}(w) = \frac{\gamma_{ss^\prime}
n_{s^\prime} m_{s^\prime}^{3/2} }{2^{3/2} T_{s^\prime}^{3/2}}
\Bigg ( \frac{\Phi( \overline{w}_{s^\prime}
)}{\overline{w}_{s^\prime}^5} -
\frac{1}{\overline{w}_{s^\prime}^4}
\frac{d\Phi}{d\overline{w}_{s^\prime}} \Bigg )
\end{equation}
is the energy scattering collision frequency. We have used the
abbreviated notation $\overline{w}_{s^\prime} = w/\sqrt{2
T_{s^\prime}/m_{s^\prime}}$, $\Phi (\overline{w}_{s^\prime}) =
(2/\sqrt{\pi}) \int_0^{\overline{w}_{s^\prime}} dy\, \exp (-y^2)$
and $d\Phi/d\overline{w}_{s^\prime} = (2/\sqrt{\pi}) \exp
(-\overline{w}_{s^\prime}^2)$.

\section{Derivation of the local $\delta f$ kinetic equation \eq{kinetickspace}} \label{app:kspace}

In this Appendix we derive \eq{kinetickspace} from \eq{kinetic}.
We Fourier analyze the equation term by term.

We change from the spatial coordinates $\bX$ to the new
coordinates $\psi (\bX)$, $\alpha(\bX, t)$ and $\theta(\bX)$.
After this coordinate transformation, and after Fourier
transforming, the term $\partial \bar h_s /\partial t + R
\Omega_\zeta \zun \cdot \nabla_\bX \bar h_s$ becomes $\partial h_s
/\partial t + [ (\partial \alpha/\partial t) + R \Omega_\zeta \zun
\cdot \nabla \alpha ] i k_\alpha h_s$. We drop the subscript
$_\bX$ in $\nabla_\bX$ for simplicity. Employing $\partial
\alpha/\partial t + R \Omega_\zeta \zun \cdot \nabla \alpha = 0$,
$\partial \bar h_s /\partial t + R \Omega_\zeta \zun \cdot
\nabla_\bX \bar h_s$ becomes $\partial h_s/\partial t$. Similarly,
the Fourier transform of $\partial \langle \bar \chi \rangle_s
/\partial t + R \Omega_\zeta \zun \cdot \nabla_\bX \langle \bar
\chi \rangle_s$ is $\partial \langle \chi \rangle_s/\partial t$.

The Fourier transform of $\bv_{d,s} \cdot \nabla_\bX \bar h_s$ is
$i k_\psi h_s \bv_{d,s} \cdot \nabla \psi + i k_\alpha h_s
\bv_{d,s} \cdot \nabla \alpha + \bv_{d,s} \cdot \nabla \theta
(\partial h_s/\partial \theta)$. The contribution $\bv_{d,s} \cdot
\nabla \theta (\partial/\partial \theta)$ is negligible since we
assume that the parallel derivative $\partial/\partial \theta$ is
small compared to $k_\bot$. Before evaluating $\bv_{d,s} \cdot
\nabla \psi$ and $\bv_{d,s} \cdot \nabla \alpha$, we rewrite the
curvature drift by employing the MHD equilibrium result $\bun
\cdot \nabla \bun = B^{-1} \nabla_\bot B + (4 \pi/B^2) (\partial
p/\partial \psi |_R) \nabla \psi$. Then, defining $v_{d,s}^\psi =
\bv_{d,s} \cdot \nabla \psi$ and $v_{d,s}^\alpha = \bv_{d,s} \cdot
\nabla \alpha$, and employing that the gradient of the
axisymmetric quantities $\Phi_0$, $B$ and $R$ can be written as
$\nabla_\bX = \nabla \psi (\partial/\partial \psi) + \nabla \theta
(\partial/\partial \theta)$, we find expressions \eq{vdpsikspace}
and \eq{vdalphakspace}. To obtain the final form of these
equations we have used $(\nabla \theta \times \zun) \cdot \nabla
\psi = R \bB \cdot \nabla \theta$ for the $\psi$ component of
$\bv_{co,s}$, $\bun \times \nabla \psi = I \bun - R B \zun$ for
the $\psi$ component of $\bv_{E0}$, $\bv_{M,s}$ and $\bv_{cf,s}$,
and $(\nabla \alpha \times \nabla \psi) \cdot \bun = B$ for the
$\alpha$ component of $\bv_{E0}$, $\bv_{M,s}$ and $\bv_{cf,s}$.

The nonlinear term $\{ \langle \chi \rangle_s, h_s \}$ is the
Fourier transform of $\bar \bv_{\chi, s} \cdot \nabla_\bX \bar
h_s$. The generalized potential $\langle \bar \chi \rangle_s$ in
\eq{chi} can be Fourier transformed by realizing that
\begin{equation} \label{chiaux}
\langle \bar \chi \rangle_s = \sum_{k_\psi, k_\alpha} \langle \chi
\exp(i S (\bx)) \rangle_s,
\end{equation}
where $\chi = \phi - c^{-1} (w_{||} A_{||} + \bw_\bot \cdot
\bA_\bot)$ and $S (\bx) = k_\psi \psi (\bx) + k_\alpha
\alpha(\bx)$. Using $S(\bx) = S(\bX + \rhobf_s) \simeq S(\bX) +
\bk_\bot \cdot \rhobf_s$, with $\bk_\bot = k_\psi \nabla \psi +
k_\alpha \nabla \alpha$, we find that the Fourier transform of
\eq{chiaux} is
\begin{eqnarray}
\langle \chi \rangle_s = \frac{1}{2\pi} \Bigg ( \phi - \frac{1}{c}
w_{||} A_{||} \Bigg ) \int d\varphi\, \exp(i \bk_\bot \cdot
\rhobf_s ) \nonumber\\ - \frac{1}{2\pi c} \bA_\bot \cdot \int
d\varphi\, \bw_\bot \exp(i \bk_\bot \cdot \rhobf_s ).
\end{eqnarray}
To obtain the final result in \eq{chikspace}, we need to employ
$B_{||} = i \bun \cdot (\bk_\bot \times \bA_\bot)$,
\begin{equation} \label{dvarphiexp}
\int d\varphi\, \exp(i\bk_\bot \cdot \rhobf_s) = 2\pi J_0(z_s)
\end{equation}
and
\begin{equation} \label{dvarphiexpwperp}
\int d\varphi\, \bw_\bot \exp (i\bk_\bot \cdot \rhobf_s) = - 2 \pi
\frac{i m_s c \mu}{Z_s e} \frac{2 J_1 (z_s)}{z_s} \bun \times
\bk_\bot,
\end{equation}
where $z_s$ is defined in \eq{defz}. To prove \eq{dvarphiexp}, we
use \eq{rho} to write $\bk_\bot \cdot \rhobf_s = - (\sqrt{2 \mu
B}/\Omega_s |\nabla \psi|) [ (k_ \psi |\nabla \psi|^2 + k_\alpha
\nabla \alpha \cdot \nabla \psi) \sin \varphi + B k_\alpha \cos
\varphi ]$. Then,
\begin{eqnarray} \label{krhosin}
\bk_\bot \cdot \rhobf_s = - z_s \sin (\varphi + \varphi_\bk),
\end{eqnarray}
with $\varphi_\bk = \arctan [ B k_\alpha /(k_ \psi |\nabla \psi|^2
+ k_\alpha \nabla \alpha \cdot \nabla \psi)]$. Using this
expression and
\begin{equation} \label{bessel}
\exp ( iz \sin \varphi ) = \sum_{n = -\infty}^\infty J_n (z) \exp
( i n\varphi),
\end{equation}
we obtain \eq{dvarphiexp}. Equation \eq{dvarphiexpwperp} is
obtained by realizing that $\bw_\bot \exp (i\bk_\bot \cdot
\rhobf_s) = i \Omega_s \bun \times \nabla_{\bk_\bot} \exp (
i\bk_\bot \cdot \rhobf_s)$ and hence $\int d\varphi\, \bw_\bot
\exp (i\bk_\bot \cdot \rhobf_s) = 2\pi i \Omega_s \bun \times
\nabla_{\bk_\bot} J_0 (z_s)$.

Finally, the last term in the right side of \eq{kinetickspace} is
the Fourier transform of the last term in \eq{kinetic}. To obtain
the final result in \eq{kinetickspace}, we use \eq{neq} to find
\begin{equation}
\frac{1}{n_s} \frac{\partial n_s}{\partial \theta} + \frac{Z_s
e}{T_s} \frac{\partial \Phi_0}{\partial \theta} - \frac{m_s R
\Omega_\zeta^2}{T_s} \frac{\partial R}{\partial \theta} = 0.
\end{equation}

\section{Gyroaveraged collision operator} \label{app:collisionave}

In this Appendix we Fourier transform the gyroaverage of the
collision operator given in Appendix~\ref{app:collision}. To
obtain the Fourier transform, we need to express the collision
operator as
\begin{equation} \label{Csskapp}
\left \langle \bar C_{ss^\prime}^{(\ell)} \right \rangle_s =
\sum_{k_\psi, k_\alpha} \left \langle C_{ss^\prime}^{(\ell)}
\right \rangle_s \exp ( iS (\bX )),
\end{equation}
where $S(\bX) = k_\psi \psi (\bX) + k_\alpha \alpha (\bX)$ depends
on the gyrocenter position $\bX$. This Appendix contains the
procedures that we use to write the different terms in the
gyroaveraged collision operator as in \eq{Csskapp}. Once the
operator is expressed in this way, it is easy to see that the
coefficients $\langle C_{ss^\prime}^{(\ell)} \rangle_s$ are the
ones that appear in \eq{kinetickspace}.

To differentiate with respect to $\bw$ in $\nabla_w \cdot
\bar{\Gammabf}_{ss^\prime}^\lambda$ and $\nabla_w \cdot
\bar{\Gammabf}_{ss^\prime}^\varepsilon$, we use that
\begin{equation} \label{hskapp}
\bar h_s = \sum_{k_\psi, k_\alpha} h_s \exp ( iS (\bX )).
\end{equation}
Employing $\nabla_w w_{||} = \bun$, $\nabla_w \mu = \bw_\bot/B$
and $\nabla_w S(\bX) = \nabla_w \bX \cdot \bk_\bot = \Omega_s^{-1}
\bun \times \bk_\bot$, we find that the Fourier transforms of the
gyroaveraged divergences of the vectors
$\bar{\Gammabf}_{ss^\prime}^\lambda$ and
$\bar{\Gammabf}_{ss^\prime}^\varepsilon$ are \cite{abel08}
\begin{equation} \label{divGammalambda}
\langle \nabla_w \cdot \Gammabf_{ss^\prime}^\lambda \rangle_s =
\frac{\partial \Gamma_{ss^\prime, w_{||}}^\lambda}{\partial
w_{||}} + \frac{\partial \Gamma_{ss^\prime, \mu}^\lambda}{\partial
\mu} - \frac{ k_\bot^2 \mu B}{\Omega_s^2} D_{ss^\prime}^\lambda
\end{equation}
and
\begin{equation} \label{divGammavarepsilon}
\langle \nabla_w \cdot \Gammabf_{ss^\prime}^\varepsilon \rangle_s
= \frac{\partial \Gamma_{ss^\prime, w_{||}}^\varepsilon}{\partial
w_{||}} + \frac{\partial \Gamma_{ss^\prime,
\mu}^\varepsilon}{\partial \mu} - \frac{ k_\bot^2 \mu
B}{\Omega_s^2} D_{ss^\prime}^\varepsilon.
\end{equation}
Here the pitch-angle scattering terms are
\begin{equation} \label{Gammalambdawpar}
\Gamma_{ss^\prime, w_{||}}^\lambda = 2 \mu \nu_{ss^\prime}^\lambda
\Bigg ( B \frac{\partial h_s}{\partial w_{||}} - w_{||}
\frac{\partial h_s}{\partial \mu} \Bigg ),
\end{equation}
\begin{equation} \label{Gammalambdamu}
\Gamma_{ss^\prime, \mu}^\lambda = 2 \mu \nu_{ss^\prime}^\lambda
\Bigg ( - w_{||} \frac{\partial h_s}{\partial w_{||}} +
\frac{w_{||}^2}{B} \frac{\partial h_s}{\partial \mu} \Bigg )
\end{equation}
and $D_{ss^\prime}^\lambda = [ 1 + (w_{||}^2/\mu B) ]
\nu_{ss^\prime}^\lambda h_s$, and the energy scattering terms are
\begin{equation} \label{Gammavarepsilonwpar}
\Gamma_{ss^\prime, w_{||}}^\varepsilon = w_{||}
\nu_{ss^\prime}^\varepsilon \Bigg ( w_{||} \frac{\partial
h_s}{\partial w_{||}} + 2\mu \frac{\partial h_s}{\partial \mu} +
\frac{m_s w^2}{T_{s^\prime}} h_s \Bigg ),
\end{equation}
\begin{equation} \label{Gammavarepsilonmu}
\Gamma_{ss^\prime, \mu}^\varepsilon = 2 \mu
\nu_{ss^\prime}^\varepsilon \Bigg ( w_{||} \frac{\partial
h_s}{\partial w_{||}} + 2 \mu \frac{\partial h_s}{\partial \mu} +
\frac{m_s w^2}{T_{s^\prime}} h_s \Bigg )
\end{equation}
and $D_{ss^\prime}^\varepsilon = \nu_{ss^\prime}^\varepsilon h_s$.
Note that $w^2 = w_{||}^2 + 2\mu B$.

To Fourier transform the gyroaveraged divergence of
$\bar{\Gammabf}_{ss^\prime}^R$, it is convenient to write the
Rosenbluth potentials $\bar H_{s^\prime} (\bx, w_{||}, \mu,
\varphi, t)$ and $\bar L_{s^\prime} (\bx, w_{||}, \mu, \varphi,
t)$ as functions of the position of the particle $\bx$ instead of
the guiding center $\bX$. To obtain $\bar H_{s^\prime}$ and $\bar
L_{s^\prime}$ from equations $\nabla_w^2 \bar L_{s^\prime} = -
8\pi \bar h_{s^\prime}$ and $\nabla_w^2 \bar H_{s^\prime} = \bar
L_{s^\prime}$, we use \eq{hskapp}, where $S (\bX) = S (\bx -
\rhobf_s) \simeq S(\bx) - \bk_\bot \cdot \rhobf_s$, and we employ
\eq{krhosin} and \eq{bessel} to write
\begin{equation}
\bar h_s = \sum_{k_\psi, k_\alpha} \sum_{n = - \infty}^\infty J_n
(z_s) h_s \exp ( i n (\varphi + \varphi_\bk) + iS(\bx) ).
\end{equation}
Using the decompositions
\begin{equation} \label{rosenLaux}
\bar L_{s^\prime} = \sum_{k_\psi, k_\alpha} \sum_{n = -
\infty}^\infty L_{s^\prime, n} \exp ( i n (\varphi + \varphi_\bk)
+ i S(\bx) )
\end{equation}
and
\begin{equation} \label{rosenHaux}
\bar H_{s^\prime} = \sum_{k_\psi, k_\alpha} \sum_{n = -
\infty}^\infty H_{s^\prime, n} \exp ( i n (\varphi + \varphi_\bk)
+ i S(\bx) ),
\end{equation}
we obtain the functions $L_{s^\prime, n} (k_\psi, k_\alpha,
w_{||}, \mu, t)$ and $H_{s^\prime, n} (k_\psi, k_\alpha, w_{||},
\mu, t)$ from equations
\begin{equation} \label{rosenLn}
\frac{\partial^2 L_{s^\prime, n}}{\partial w_{||}^2} + \frac{2}{B}
\frac{\partial}{\partial \mu} \left ( \mu \frac{\partial
L_{s^\prime, n}}{\partial \mu} \right ) - \frac{n^2}{2\mu B}
L_{s^\prime, n} = - 8 \pi J_n (z_{s^\prime}) h_{s^\prime}
\end{equation}
and
\begin{equation} \label{rosenHn}
\frac{\partial^2 H_{s^\prime, n}}{\partial w_{||}^2} + \frac{2}{B}
\frac{\partial}{\partial \mu} \left ( \mu \frac{\partial
H_{s^\prime, n}}{\partial \mu} \right ) - \frac{n^2}{2\mu B}
H_{s^\prime, n} = L_{s^\prime, n}.
\end{equation}
With the decompositions \eq{rosenLaux} and \eq{rosenHaux}, and
$\nabla_w w_{||} = \bun$, $\nabla_w \mu = \bw_\bot/B$, $\nabla_w
\varphi = - \bw \times \bun/2\mu B$, $\nabla_w \nabla_w w_{||} =
0$, $\nabla_w \nabla_w \mu = (\matI - \bun \bun)/B$ and $\nabla_w
\nabla_w \varphi = [ \bw_\bot (\bw \times \bun) + (\bw \times
\bun) \bw_\bot ]/ 4\mu^2 B^2$, we obtain
\begin{eqnarray}
\nabla_w \cdot \bar{\Gammabf}_{ss^\prime}^R = \sum_{k_\psi,
k_\alpha} \sum_{n= -\infty}^\infty \Bigg [ \frac{\partial
G_{ss^\prime, w_{||}, n}^R}{\partial w_{||}} + \frac{\partial
G_{ss^\prime, \mu, n}^R}{\partial \mu} \nonumber\\ -
\frac{k_\bot^2 \mu B}{\Omega_s^2} \Delta_{ss^\prime,n}^R \Bigg ]
\exp ( i n(\varphi + \varphi_\bk) + i S(\bx) ),
\end{eqnarray}
where
\begin{eqnarray}
G_{ss^\prime, w_{||}, n}^R = - \frac{m_s \gamma_{ss^\prime}
f_{Ms}}{T_s} \Bigg ( w_{||} \frac{\partial^2 H_{s^\prime,
n}}{\partial w_{||}^2} \nonumber\\ + 2\mu \frac{\partial^2
H_{s^\prime, n}}{\partial w_{||} \partial \mu} +
\frac{T_s}{m_{s^\prime}} \frac{\partial L_{s^\prime, n}}{\partial
w_{||}} \Bigg ),
\end{eqnarray}
\begin{eqnarray}
G_{ss^\prime, \mu, n}^R = - \frac{2\mu m_s \gamma_{ss^\prime}
f_{Ms}}{B T_s} \Bigg ( w_{||} \frac{\partial^2 H_{s^\prime,
n}}{\partial w_{||} \partial \mu} \nonumber\\ + 2\mu
\frac{\partial^2 H_{s^\prime, n}}{\partial \mu^2} + \frac{\partial
H_{s^\prime, n}}{\partial \mu} + \frac{T_s}{m_{s^\prime}}
\frac{\partial L_{s^\prime, n}}{\partial \mu} \Bigg )
\end{eqnarray}
and
\begin{eqnarray}
\Delta_{ss^\prime, n}^R = - \frac{n^2}{z_s^2} \frac{m_s
\gamma_{ss^\prime} f_{Ms}}{\mu B T_s} \Bigg ( w_{||}
\frac{\partial H_{s^\prime, n}}{\partial w_{||}} \nonumber\\ + 2
\mu \frac{\partial H_{s^\prime, n}}{\partial \mu} - H_{s^\prime,
n} + \frac{T_s}{m_{s^\prime}} L_{s^\prime, n} \Bigg ).
\end{eqnarray}
Using $S(\bx) = S(\bX + \rhobf_s) \simeq S(\bX) + \bk_\bot \cdot
\rhobf_s$ and \eq{bessel}, we find
\begin{equation}
\exp (i S(\bx)) = \sum_{n = -\infty}^\infty J_n (z_s) \exp ( - i
n(\varphi + \varphi_\bk) + i S(\bX)),
\end{equation}
giving that the Fourier transform of the gyroaverage of $\nabla_w
\cdot \bar{\Gammabf}_{ss^\prime}^R$ is
\begin{eqnarray} \label{divGammaR}
\langle \nabla_w \cdot \Gammabf_{ss^\prime}^R \rangle_s = \sum_{n=
-\infty}^\infty  J_n (z_s) \Bigg ( \frac{\partial G_{ss^\prime,
w_{||}, n}^R}{\partial w_{||}} \nonumber\\ + \frac{\partial
G_{ss^\prime, \mu, n}^R}{\partial \mu} - \frac{k_\bot^2 \mu
B}{\Omega_s^2} \Delta_{ss^\prime, n}^R \Bigg ).
\end{eqnarray}
This equation combined with \eq{divGammalambda} and
\eq{divGammavarepsilon} gives that the Fourier transform of the
collision operator is
\begin{eqnarray} \label{collisionkspace}
\left \langle C_{ss^\prime}^{(\ell)} \right \rangle_s =
\frac{\partial}{\partial w_{||}} \left ( \Gamma_{ss^\prime,
w_{||}}^\lambda + \Gamma_{ss^\prime, w_{||}}^\varepsilon +
\Gamma_{ss^\prime, w_{||}}^R \right ) \nonumber\\ +
\frac{\partial}{\partial \mu} \left ( \Gamma_{ss^\prime,
\mu}^\lambda + \Gamma_{ss^\prime, \mu}^\varepsilon +
\Gamma_{ss^\prime, \mu}^R \right ) \nonumber\\ - \frac{ k_\bot^2
\mu B}{\Omega_s^2} \left ( D_{ss^\prime}^\lambda +
D_{ss^\prime}^\varepsilon + D_{ss^\prime}^R \right ),
\end{eqnarray}
where
\begin{equation}
\Gamma_{ss^\prime, w_{||}}^R = \sum_{n = -\infty}^\infty J_n (z_s)
G_{ss^\prime, w_{||}, n}^R,
\end{equation}
\begin{equation}
\Gamma_{ss^\prime, \mu}^R = \sum_{n = -\infty}^\infty J_n (z_s)
G_{ss^\prime, \mu, n}^R
\end{equation}
and
\begin{eqnarray}
D_{ss^\prime}^R = \sum_{n = -\infty}^\infty \Bigg [ \frac{J_{n-1}
(z_s) - J_{n+1} (z_s)}{2 \mu z_s} G_{ss^\prime, \mu, n}^R
\nonumber\\ + J_n (z_s) \Delta_{ss^\prime, w_{||}, n}^R \Bigg ].
\end{eqnarray}
To obtain the final result in \eq{collisionkspace}, we integrated
the first two terms in \eq{divGammaR} by parts in $w_{||}$ and
$\mu$, and we used $\partial J_n(z_s)/\partial w_{||} = 0$ and
$\partial J_n(z_s)/\partial \mu = (k_\bot \sqrt{B}/\Omega_s
\sqrt{2\mu}) (dJ_n/dz_s)$, where $dJ_n/dz_s = [J_{n-1} (z_s) -
J_{n+1} (z_s)]/2$.

\section{Derivation of the local $\delta f$ equations for the electromagnetic fields} \label{app:kspace2}

In this Appendix we show how to obtain
\eq{qnkspace}-\eq{perppresskspace} from \eq{qn}-\eq{perppress}. We
only need to Fourier transform \eq{qn}-\eq{perppress}. In equation
\eq{qn}, the integral $\int dw_{||}\, d\mu\, d\varphi\, \bar h_s$
can be written as
\begin{equation}
\sum_{k_\psi, k_\alpha} \int dw_{||}\, d\mu\, h_s \int d\varphi\,
\exp (iS(\bX)),
\end{equation}
where $S(\bX) = k_\psi \psi(\bX) + k_\alpha \alpha (\bX)$. Using
$S(\bX) = S(\bx - \rhobf_s) \simeq S(\bx) - \bk_\bot \cdot
\rhobf_s$ and \eq{dvarphiexp}, we find $\int d\varphi\, \exp
(iS(\bX)) = 2\pi \exp(iS(\bx)) J_0(z_s)$, and \eq{qnkspace}
immediately follows. Equations \eq{parcurrkspace} and
\eq{perppresskspace} are obtained from \eq{parcurr} and
\eq{perppress} in a similar way. In equation \eq{perppress} we
find the integral $\int d\varphi\, \bw_\bot \exp (iS(\bX))$. Using
$S(\bX) \simeq S(\bx) - i\bk_\bot \cdot \rhobf_s$ and
\eq{dvarphiexpwperp}, we obtain \eq{perppresskspace}.

\section{Derivation of the local $\delta f$ flux of toroidal angular momentum} \label{app:flux}

In this Appendix we show how to obtain
\eq{piparphikspace}-\eq{piBperpkspace} from \eq{pipar}-\eq{g}. We
only show in detail how to obtain \eq{piparphikspace} from the
term proportional to $-(c/B) \nabla \bar \phi \times \bun$ in
\eq{pipar}. We then briefly cover the other terms. Writing the
flux surface average explicitly, we obtain
\begin{eqnarray}
\pi_{s,||}^\phi = \frac{m_s c I}{V^\prime \Delta \psi}
\sum_{k_\psi, k_\alpha, k^\prime_\psi, k^\prime_\alpha} \int
d\psi\, d\theta\, d\zeta\, \frac{\mathcal{J}}{B} \nonumber \\
\times ik_\alpha \phi(k_\psi, k_\alpha) \int d^3w\, h_s
(k_\psi^\prime, k_\alpha^\prime) w_{||} \nonumber\\ \times \exp (i
\sigma (\bx, \bX, k_\psi, k_\alpha, k_\psi^\prime, k_\alpha^\prime
) ),
\end{eqnarray}
where we employ the abbreviated notation $\sigma (\bx, \bX,
k_\psi, k_\alpha, k_\psi^\prime, k_\alpha^\prime) = k_\psi
\psi(\bx) + k_\alpha \alpha (\bx) + k_\psi^\prime \psi (\bX) +
k_\alpha^\prime \alpha (\bX)$. Using $\bX = \bx - \rhobf_s$,
$\sigma$ can be simplified to $\sigma \simeq (k_\psi +
k_\psi^\prime) \psi(\bx) + (k_\alpha + k_\alpha^\prime) \alpha
(\bx) - i \bk_\bot^\prime \cdot \rhobf_s$, with $\bk_\bot^\prime =
k_\psi^\prime \nabla \psi + k_\alpha^\prime \nabla \alpha$. Using
$\alpha = \zeta - q \vartheta - \Omega_\zeta t$, and noting that
the integral over $\psi$ and $\zeta$ is over distances much longer
than the perpendicular correlation length, we find that only the
contributions with $k_\psi^\prime = - k_\psi$ and $k_\alpha^\prime
= - k_\alpha$ do not vanish, leading to
\begin{eqnarray}
\pi_{s,||}^\phi = \frac{2 \pi i m_s c I}{V^\prime} \sum_{k_\psi,
k_\alpha} k_\alpha \int d\theta\, \frac{\mathcal{J}}{B}
\phi(k_\psi, k_\alpha) \nonumber \\ \times \int d^3w\, h_s (-
k_\psi, - k_\alpha) w_{||} \exp (i \bk_\bot \cdot \rhobf_s ).
\end{eqnarray}
To obtain the final result in \eq{piparphikspace}, we use $d^3w =
B\, dw_{||}\, d\mu\, d\varphi$ and \eq{dvarphiexp}.

Equations \eq{piparAparkspace}-\eq{piBperpkspace} are obtained in
a similar way. To obtain the final expressions, we only need
\eq{dvarphiexp}, \eq{dvarphiexpwperp} and
\begin{eqnarray} \label{wwbexp}
\int d\varphi\, \bw_\bot (\bw \times \bun) \exp(i\bk_\bot \cdot
\rhobf_s) \nonumber\\ = 2\pi  \mu B \bigg [ \frac{2 J_1(z_s)}{z_s}
\bun \times \matI - \frac{\mu B}{2\Omega_s^2} G (z_s) (\bun \times
\bk_\bot) \bk_\bot \bigg ],
\end{eqnarray}
where $G(z_s) = [8J_1 (z_s) + 4z_s J_2 (z_s) - 4 z_s
J_0(z_s)]/z_s^3$. This result is obtained by realizing that
$\bw_\bot (\bw \times \bun) \exp(i\bk_\bot \cdot \rhobf_s) = -
\Omega_s^2 \bun \times \nabla_{\bk_\bot} \nabla_{\bk_\bot} \exp (i
\bk_\bot \cdot \rhobf_s)$, giving $\int d\varphi \, \bw_\bot (\bw
\times \bun) \exp(i\bk_\bot \cdot \rhobf_s) = - 2\pi \Omega_s^2
\bun \times \nabla_{\bk_\bot} \nabla_{\bk_\bot} J_0 (z_s)$.

\section{Up-down reflection of the equilibrium} \label{app:updown}

In this Appendix we show that it is possible to find a poloidal
angle definition $\theta$ for the up-down reflection of an
equilibrium such that Table~\ref{table_geom} is satisfied.

We first calculate the geometrical coefficients $B$, $\bB \cdot
\nabla \theta$, $\mathcal{J}$, $|\nabla \psi|^2$, $\nabla \psi
\cdot \nabla \theta$ and $|\nabla \theta|^2$ as functions of
$\psi$ and $\theta$ for some choice of poloidal angle definition
$\theta$. We invert the functions $\psi (R, Z)$ and $\theta (R,
Z)$, where $R$ and $Z$ are the typical radial and axial
coordinates satisfying $\nabla R \times \nabla Z = \zun$. We then
write all the coefficients using the functions $R (\psi, \theta)$
and $Z(\psi, \theta)$. The poloidal component of the magnetic
field $\bB \cdot \nabla \theta = (\nabla \zeta \times \nabla \psi)
\cdot \nabla \theta$ is written in terms of $R$ and $Z$ as
\begin{equation} \label{Btheta}
\bB \cdot \nabla \theta = \left [ R \left ( \frac{\partial
R}{\partial \psi} \frac{\partial Z}{\partial \theta} -
\frac{\partial Z}{\partial \psi} \frac{\partial R}{\partial
\theta} \right ) \right ]^{-1}.
\end{equation}
The Jacobian $\mathcal{J} = |\bB \cdot \nabla \theta|^{-1}$. The
products $|\nabla \psi|^2$, $\nabla \psi \cdot \nabla \theta$ and
$|\nabla \theta|^2$ are
\begin{equation} \label{gradpsi2}
|\nabla \psi|^2 = \left ( \frac{R}{\mathcal{J}} \right )^2 \left [
\left ( \frac{\partial R}{\partial \theta} \right )^2 + \left
(\frac{\partial Z}{\partial \theta} \right )^2 \right ],
\end{equation}
\begin{equation} \label{gradpsigradtheta}
\nabla \psi \cdot \nabla \theta = - \left (\frac{R}{\mathcal{J}}
\right )^2 \left (\frac{\partial R}{\partial \psi} \frac{\partial
R}{\partial \theta} + \frac{\partial Z}{\partial \psi}
\frac{\partial Z}{\partial \theta} \right )
\end{equation}
and
\begin{equation} \label{gradtheta2}
|\nabla \theta|^2 = \left ( \frac{R}{\mathcal{J}} \right )^2 \left
[ \left ( \frac{\partial R}{\partial \psi} \right )^2 + \left (
\frac{\partial Z}{\partial \psi} \right )^2 \right ].
\end{equation}
The magnitude of the magnetic field is $B^2 = ( I^2 + |\nabla
\psi|^2)/R^2$.

For an equilibrium given by $R(\psi, \theta)$ and $Z(\psi,
\theta)$, the equilibrium described by $R(\psi, \theta)$ and $-
Z(\psi, \theta)$ is the up-down reflection. This is a way of
choosing the poloidal angle definition $\theta$ for the up-down
reflection. With this choice, equivalent to replacing $Z$ by $-Z$
in \eq{Btheta}-\eq{gradtheta2}, we obtain most of the results in
Table~\ref{table_geom}. The poloidally varying quantities $n_s$,
$p$ and $\partial p/\partial \psi|_R$ remain the same because they
depend on $\theta$ only through the function $R(\psi, \theta)$,
which is unchanged upon up-down reflection.

\section{Proof of the results in Table~\ref{table_geom_2}} \label{app:updown2}

In this Appendix we evaluate the coefficients in
Table~\ref{table_geom_2} and express them as functions of other
coefficients whose behavior under the transformations in
Table~\ref{table_change} we know.

Using $\nabla \alpha \times \bB = ( \nabla \psi \cdot \nabla
\alpha ) \nabla \alpha - |\nabla \alpha|^2 \nabla \psi$, we find
\begin{equation}
(\nabla \alpha \times \bB) \cdot \nabla \theta = (\nabla \alpha
\cdot \nabla \psi)(\nabla \alpha \cdot \nabla \theta) - |\nabla
\alpha|^2 (\nabla \psi \cdot \nabla \theta).
\end{equation}
Using $\nabla R \times \zun = - \nabla Z$, we find
\begin{equation}
(\nabla R \times \zun) \cdot \nabla \alpha = - \frac{\partial
Z}{\partial \psi} \nabla \psi \cdot \nabla \alpha - \frac{\partial
Z}{\partial \theta} \nabla \theta \cdot \nabla \alpha.
\end{equation}
Finally, the coefficients $|\nabla \alpha|^2$, $\nabla \alpha
\cdot \nabla \psi$ and $\nabla \alpha \cdot \nabla \theta$ are
\begin{eqnarray}
|\nabla \alpha|^2 = \frac{1}{R^2} + \left ( \vartheta
\frac{\partial q}{\partial \psi} + q \frac{\partial
\vartheta}{\partial \psi} + t \frac{\partial
\Omega_\zeta}{\partial \psi} \right )^2 |\nabla \psi|^2
\nonumber\\ + \frac{2I}{R^2 \bB \cdot \nabla \theta} \left (
\vartheta \frac{\partial q}{\partial \psi} + q \frac{\partial
\vartheta}{\partial \psi} + t \frac{\partial
\Omega_\zeta}{\partial \psi} \right ) \nabla \psi \cdot \nabla
\theta \nonumber\\ + \left ( \frac{I}{R^2 \bB \cdot \nabla \theta}
\right )^2 |\nabla \theta|^2,
\end{eqnarray}
\begin{eqnarray}
\nabla \alpha \cdot \nabla \psi = - \left ( \vartheta
\frac{\partial q}{\partial \psi} + q \frac{\partial
\vartheta}{\partial \psi} + t \frac{\partial
\Omega_\zeta}{\partial \psi} \right ) |\nabla \psi|^2 \nonumber\\
- \frac{I}{R^2 \bB \cdot \nabla \theta} \nabla \psi \cdot \nabla
\theta
\end{eqnarray}
and
\begin{eqnarray}
\nabla \alpha \cdot \nabla \theta = - \left ( \vartheta
\frac{\partial q}{\partial \psi} + q \frac{\partial
\vartheta}{\partial \psi} + t \frac{\partial
\Omega_\zeta}{\partial \psi} \right ) \nabla \psi \cdot \nabla
\theta \nonumber \\ - \frac{I}{R^2 \bB \cdot \nabla \theta}
|\nabla \theta|^2,
\end{eqnarray}
with the function $\vartheta(\psi, \theta)$ defined in
\eq{vartheta}.

With these expressions, and employing Table~\ref{table_geom}, we
can show that transformation I, which is an up-down reflection and
a sign change in $\Omega_\zeta$ and $\partial
\Omega_\zeta/\partial \psi$, transforms the coefficients as in
Table~\ref{table_geom_2}. We can prove the same for transformation
II, which is an up-down reflection and a sign change in $I$.

\end{document}